\documentclass[twocolumn,trackchanges]{aastex63}
\pdfoutput=1

\accepted{June 2, 2021}
\submitjournal{\apj}

\shorttitle{Classification of Filament Formation Mechanisms in Magnetized Molecular Clouds}
\shortauthors{Abe et al.}

\begin{document}

\title{Classification of Filament Formation Mechanisms in Magnetized Molecular Clouds}

\email{d.abe@nagoya-u.jp}

\author[0000-0001-6891-2995]{Daisei Abe}
\affiliation{Department of Physics, Graduate School of Science, Nagoya University, Furo-cho, Chikusa-ku, Nagoya 464-8602, Japan}


\author[0000-0002-7935-8771]{Tsuyoshi Inoue}
\affiliation{Department of Physics, Graduate School of Science, Nagoya University, Furo-cho, Chikusa-ku, Nagoya 464-8602, Japan}

\author[0000-0003-4366-6518]{Shu-ichiro Inutsuka}
\affiliation{Department of Physics, Graduate School of Science, Nagoya University, Furo-cho, Chikusa-ku, Nagoya 464-8602, Japan}

\author[0000-0002-8125-4509]{Tomoaki Matsumoto}
\affiliation{Faculty of Sustainability Studies, Hosei University, Fujimi, Chiyoda-ku, Tokyo 102-8160, Japan}








\begin{abstract}
Recent observations of molecular clouds show that dense filaments are the sites of present-day star formation.
Thus, it is necessary to understand the filament formation process because these filaments provide the initial condition for star formation.
Theoretical research suggests that shock waves in molecular clouds trigger filament formation.
Since several different mechanisms have been proposed for filament formation, the formation mechanism of the observed star-forming filaments requires clarification.
In the present study, we perform a series of isothermal magnetohydrodynamics simulations of filament formation.
We focus on the influences of shock velocity and turbulence on the formation mechanism and identified three different mechanisms for the filament formation.
The results indicate that when the shock is fast, at shock velocity $v_{\mathrm{sh}} \simeq 7\ \mathrm{km s^{-1}}$, the gas flows driven by the curved shock wave create filaments irrespective of the presence of turbulence and self-gravity.
However, at a slow shock velocity $v_{\mathrm{sh}} \simeq 2.5\ \mathrm{km s^{-1}}$, the compressive flow component involved in the initial turbulence induces filament formation.
When both the shock velocities and turbulence are low, the self-gravity in the shock-compressed sheet becomes important for filament formation.
Moreover, we analyzed the line-mass distribution of the filaments and showed that strong shock waves can naturally create high-line-mass filaments such as those observed in the massive star-forming regions in a short time.
We conclude that the dominant filament formation mode changes with the velocity of the shock wave triggering the filament formation.

\end{abstract}


\section{Introduction} \label{sec:intro}
Stars are formed in dense regions in molecular clouds~\protect\citep[e.g.,][]{Lada2010,Enoch2007,Andreetal2014,HennebelleInutsuka2019}.
Recent observations by the \textit{Herschel} space telescope revealed that dense filamentary structures are ubiquitous in nearby molecular clouds~\protect\citep[e.g.,][]{Andre2010,Arzoumanian2011}.
Additionally, star-forming cores and young stellar objects are embedded along the filaments, which indicate their crucial role in star formation~\protect\citep{Konyves2015}.
To create stars in filaments, their line-masses have to exceed the critical line-mass for gravitational instability, $M_{\mathrm{line,cr}} = 2c_{\mathrm{s}}^2 /G \simeq 17\ \mathrm{M_{\odot}\ pc^{-1}}$, where $c_{\mathrm{s}} \simeq 0.2\ \mathrm{km\ s^{-1}}$ is the isothermal sound speed of typical molecular clouds~\protect\citep[e.g.,][]{Stodolkiewicz1963,Ostriker1964,InutsukaMiyama1992,Inutsuka1997}.

Previous theoretical researches on filament formation proposed several types of mechanisms.
The first one is a well-known self-gravitational fragmentation of a sheetlike cloud~\protect\citep{Tomisaka1983,Miyama1987a,Miyama1987b,Nagai1998,Kitsionas2007,Balfour2015,Balfour2017}, which is created when molecular clouds are shock-compressed.
Such compression of a molecular cloud naturally occurs owing to cloud-cloud collision, feedback from massive stars including supernovae and expanding HII regions, and encounters with galactic spiral shock.
In this paper, we call this type of filament formation type G formation.
According to linear stability analysis conducted by \protect\citet{Nagai1998}, gravitational instability creates filaments with line-masses larger than  but comparable to the critical line-mass when the width of the sheet is comparable to the Jeans length (i.e., the self-gravity is important in the dynamics).
\protect\citet{Padoan1999}, \protect\citet{PudritzKevlahan2013},  \protect\citet{Matsumoto2015}, and, \protect\citet{Federrath2016} showed filament formation induced by turbulence in molecular clouds by using three-dimensional magnetohydrodynamic (MHD) simulation.
\protect\citet{Padoan1999} reported that the turbulent velocity given in an initially uniform molecular gas induces the formation of shock-compressed sheets, and then the interaction of two sheets creates a filament at the intersection of them.
We call this mechanism type I filament formation.
It should be noted that the type I process occurs in the super-Alfv\'{e}nic case, in which the initial Alfv\'{e}nic Mach number $\mathcal{M}_{\mathrm{A,i}} \gtrsim 10$.
\protect\citet{Inoue2013}, \protect\citet{Vaidya2013}, \protect\citet{Inutsuka2015}, and \protect\citet{Inoue2018} reported that the filaments are generated when a shock wave sweeps a cloud containing density inhomogeneity or clumps.
In this case, a dense blob embedded in a magnetized molecular cloud is transformed into a dense filament in the shock-compressed layer owing to the effect of an oblique (or curved) MHD shock wave.
Throughout this study, this filament formation mode is called type O mode.
It should be noted that the effects of the thermal instability and turbulence cause molecular clouds to be highly clumpy by nature~\protect\citep[e.g.,][]{InoueInutsuka2012}.
Note that in most theoretical works that study structure formation in molecular clouds by turbulence, a uniform density gas is used as an initial condition.
Therefore the type O mode is missing in the studies that assume the uniform initial density.
As we show subsequently, the type O mode dominates over other modes when the shock Mach number is high.

\protect\citet{Chen2014} and \protect\citet{Chen2015} showed another filament formation behind shock waves that is different from type O.
They found that filaments are formed by converging gas flow components along the local magnetic field in the case the substantial turbulent motions are given in the initial condition.
We call this mode type C filament creation mechanism.
This mechanism creates the filaments that are perpendicular to the background magnetic field when the turbulence is sub-Alfv\'{e}nic~\protect\citep{Chen2014,Chen2015,PlanckCollaboration2016}.
This is because the component of turbulent velocity perpendicular to the magnetic field is suppressed by magnetic tension force, and only motion parallel to the magnetic field can compress the gas.
\protect\citet{Padoan1999} also reported the type C process in the trans-Alfv\'{e}nic case ($\mathcal{M}_{\mathrm{A,i}}\sim 1$).

\protect\citet{Hennebelle2013} reported that a small clump in a turbulent molecular cloud is stretched by turbulent shear flows and evolves into a small line-mass filament parallel to the magnetic field~\protect\citep[see also][for the origin of HI filament/fiber]{InoueInutsuka2016}.
We call this mode type S mechanism, and this type S creates filaments that are parallel to the magnetic field lines in contrast to the other types that create filaments perpendicular to the magnetic field lines.
\protect\citet{XuLazarian2019} showed that the anisotropic nature of turbulent MHD eddies results in filamentary structure formation parallel to a magnetic field.
We summarize the filament formation mechanism in table \ref{tab:FormationMechanism}.
\begin{table*}
\caption{Filament formation mechanisms.\label{tab:FormationMechanism}}
 \centering
  \begin{tabular}{ccl}
   \hline
   Category & Filament vs. Magnetic field & A brief description of the formation mechanism \\
   \hline \hline
    type G & perpendicular & self-gravitational fragmentation of a sheetlike cloud \\
    type I & - & the intersection of two shocked layers \\
    type O & perpendicular & the effect of an oblique MHD shock wave \\
    type C & perpendicular & converging gas flow components along the magnetic field in the post-shock layer \\
    type S & parallel & stretching by turbulent shear flows, small line-mass \\
   \hline
  \end{tabular}
  \tablecomments{Original papers of above types are following.
  Type G: \citet{Tomisaka1983, Miyama1987a, Miyama1987b, Nagai1998, Kitsionas2007, Balfour2015, Balfour2017}
  Type I: \citet{Padoan1999, PudritzKevlahan2013, Matsumoto2015, Federrath2016} 
  Type O: \citet{Inoue2013, Vaidya2013, Inoue2018}
  Type C: \citet{Chen2014, Chen2015, Padoan1999}
  Type S: \citet{Hennebelle2013, InoueInutsuka2016}
  }
\end{table*}

In this way, many filament formation mechanisms have been proposed.
However, it is still unclear which type is responsible for the creation of star-forming filaments.
Most of the proposed mechanisms (type I, O, C, and possibly type G) are triggered by shock compression.
Thus, in this study, we perform a series of isothermal MHD simulations of filament formation by shock waves, focusing on the influences of shock strength, turbulence, and self-gravity on the filament formation mechanism.
Note here that, since type I is effective only in highly super-Alfv\'enic turbulence, we hardly observe it in the results of our simulations that are performed under a realistic 10 $\mu$G initial magnetic field.
In addition, we analyze the line-mass distribution of the simulated filaments.
The paper is organized as follows:
In \S 2, we provide the setup of our simulations, and we show and interpret the results including the filament line-mass distribution in \S 3.
In \S 4, simple models are presented that account for the dominant filament formation timescale.
Finally, we summarize the results in \S 5.

\section{Setup of Simulations} \label{sec:Setup of Simulations}
\begin{figure}[ht!]
\plotone{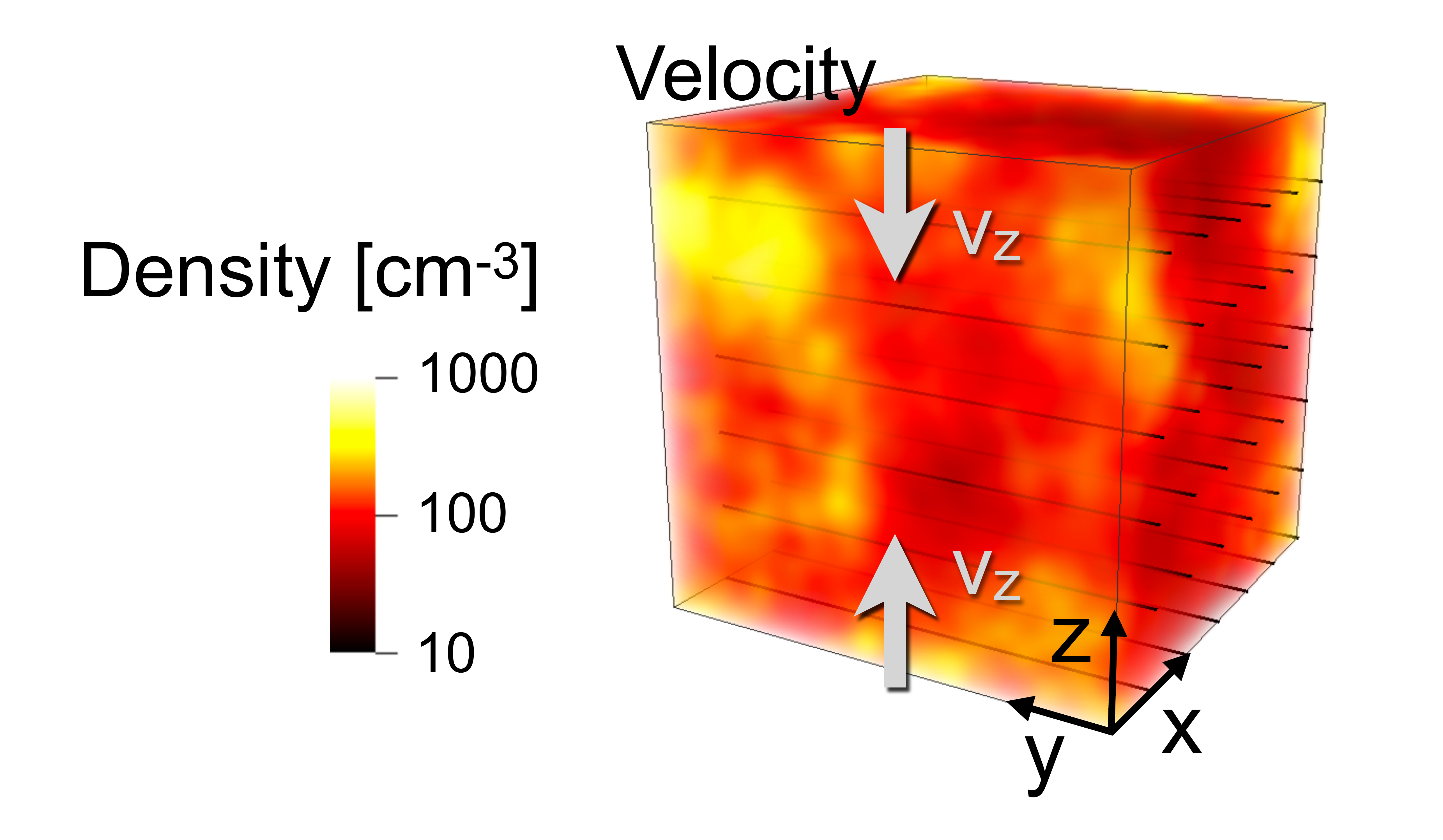}
\caption{Schematic diagram of initial condition.\label{fig:inicon}
The color bar represents the density magnitude; the black lines are the initial magnetic field lines; and the grey arrows indicate the orientations of the converging flows.
}
\end{figure}
We solve the isothermal MHD equations with self-gravity by using the SFUMATO code \protect\citep{Matsumoto2007} which integrates the MHD equations using a Godunov-type scheme with a HLLD Riemann solver \protect\citep{Miyoshi2005} having third- and second-order accuracies in space and time, respectively.
A divergence cleaning method~\protect\citep{Dedner2002} ensures the divergence free condition, $\nabla \cdot \mathbf{B} = 0$.
Poisson's equation is solved by using the multi-grid method.

\begin{table*}
\caption{Model parameters.\label{tab:modelparameters}}
 \centering
  \begin{tabular}{ccccc}
   \hline
   Model Name & Collision Velocity $v_{\mathrm{coll}}~[\mathrm{km\ s^{-1}}]$ & Shock Velocity $v_{\mathrm{sh}}~[\mathrm{km\ s^{-1}}]$ & Self-Gravity & Turbulence \\
   \hline \hline
    v12GyTn & 12 & 7.0 & Yes & No \\
    v12GnTn & 12 & 7.0 & No & No \\
    v12GyTy & 12 & 7.0 & Yes & Yes \\
    v12GnTy & 12 & 7.0 & No & Yes \\
    v10GyTn & 10 & 6.0 & Yes & No \\
    v8GyTn & 8.0 & 5.0 & Yes & No \\
    v6GyTn & 6.0 & 4.0 & Yes & No \\
    v3GyTn & 3.0 & 2.5 & Yes & No \\
    v3GnTn & 3.0 & 2.5 & No & No \\
    v3GyTy & 3.0 & 2.5 & Yes & Yes \\
    v3GnTy & 3.0 & 2.5 & No & Yes \\
   \hline
  \end{tabular}
  \tablecomments{Gas continuously flows into the calculation box from the upper and lower boundaries and initially collides at $z$~=~3~pc with the relative velocity $v_{\mathrm{coll}}$.
  After the collision the shock velocity~$v_{\mathrm{sh}}$ in the rest frame of the upstream gas becomes larger than $v_{\mathrm{coll}}/2$, and thus, the thickness of the shock-compressed layer expands.
  We show the relation between collision velocity and shock velocity in Eqs. \ref{equation:shock velocity}}.
\end{table*}

We investigate the mechanism of filament formation induced by shock waves by performing converging flow simulations.
A schematic illustration of the initial condition is shown in Figure \ref{fig:inicon}.
We prepare a cubic numerical domain of side lengths $L_{\mathrm{box}} = 6.0\ \mathrm{pc}$ consisting of $512^3$ uniform numerical cells, indicating that the physical resolution $\Delta x$ is approximately $0.012\ \mathrm{pc}$.
We initially set the isothermal gas characterized by the isothermal sound speed at $c_{\mathrm{s}}=0.2\ \mathrm{km\ s^{-1}}$.
Because molecular clouds are highly inhomogeneous by nature, we initially add isotropic density fluctuations given as a superposition of sinusoidal functions with various wavenumbers from $2 \pi /L_{\mathrm{box}} \leq |k| \leq 32 \pi /L_{\mathrm{box}}$ and random phases.
The power spectrum of the density fluctuations is given by $(\log \rho)^2_k \propto k^{-4}$ , which can be expected as a consequence of supersonic turbulence~\protect\citep{Beresnyak2005,Elmegreen2004,Scalo2004,Larson1981,Heyer2004}.
Thus, the initial density structure of our simulations is parameterized by mean density $\bar{n}_0 = \bar{\rho}_0/m = 100\ \mathrm{cm^{-3}}$ and dispersion $\Delta n/\bar{n}_0 = 0.5$, where $\bar{\rho}_0$ and $m=2.4\ m_{\mathrm{proton}}$ are the mean mass density and the mean mass of the molecular gas particles, respectively.
In addition to density fluctuations, we set the initial turbulent velocity field depending on the model summarized in Table \ref{tab:modelparameters}.
The initial turbulent velocity fluctuation has a dispersion of $1.0\ \mathrm{km\ s^{-1}}$ with a power spectrum of $v^2 _k \propto k^{-4}$ following Larson's law~\protect\citep{Larson1981}.

In addition to the turbulent component, we set the initial coherent velocity component as $v_z(z) = -(v_{\mathrm{coll}}/2) \tanh[z - 3]$, i.e., two flows colliding head-on in the x-y plane at $z=3\ \mathrm{pc}$ with a relative velocity of $v_{\mathrm{coll}}$.
In the previous studies that reported the type C filament formation, the filaments are created in the shock-compressed layer with shock velocity of a few km s$^{-1}$.
However, in the studies that found the type O mechanism, high shock velocity cases of $\sim 10\ \mathrm{km\ s^{-1}}$ are studied.
To systematically study the filament formation mechanism, we perform simulations using various shock velocities.
Specifically, we examine cases with $v_{\mathrm{coll}} = 3$, $v_{\mathrm{coll}} = 6$, $v_{\mathrm{coll}} = 8$, $v_{\mathrm{coll}} = 10$, and $12\ \mathrm{km\ s^{-1}}$.
Table \ref{tab:modelparameters} includes a summary of the model parameters.
In the model names, the number following “v" represents the value of $v_{\mathrm{coll}}$ in units of $\mathrm{km\ s^{-1}}$, and the characters “y" and “n" following “G" and “T" represent simulation with and without self-gravity and initial turbulence, respectively.

At the x-y boundary planes, the velocity is fixed at $v_{\mathrm{coll}}/2$, and the density is given by $n_0 (x, y, z = v_{\mathrm{coll}} t/2)$ for the $z = 6$ pc plane and $n_0 (x, y, z = L_{\mathrm{box}} - v_{\mathrm{coll}} t/2)$ for the $z = 0$ pc plane, where $n_0(x, y, z)$ is the initial density field including fluctuations.
For the $z$-$x$ and $y$-$z$ boundary planes, we impose periodic boundary conditions.


We initially set a uniform magnetic field of $\mathbf{B}_0 = (0,10\ \mathrm{\mu G},0)$, which is perpendicular to the direction of the shock propagation.
This magnetic field strength is consistent with observed magnitude in molecular clouds~\protect\citep[e.g., ][]{Crutcher2012,Heiles2005a}.
Because the magnetic field component perpendicular to the direction of the shock propagation is expected to be strongly amplified by the shock compression, whereas the parallel component is not, the initial z-component of the magnetic field would play a minor role even if it is given.

\begin{figure*}
\begin{center}
    \includegraphics[clip,width=17.5cm]{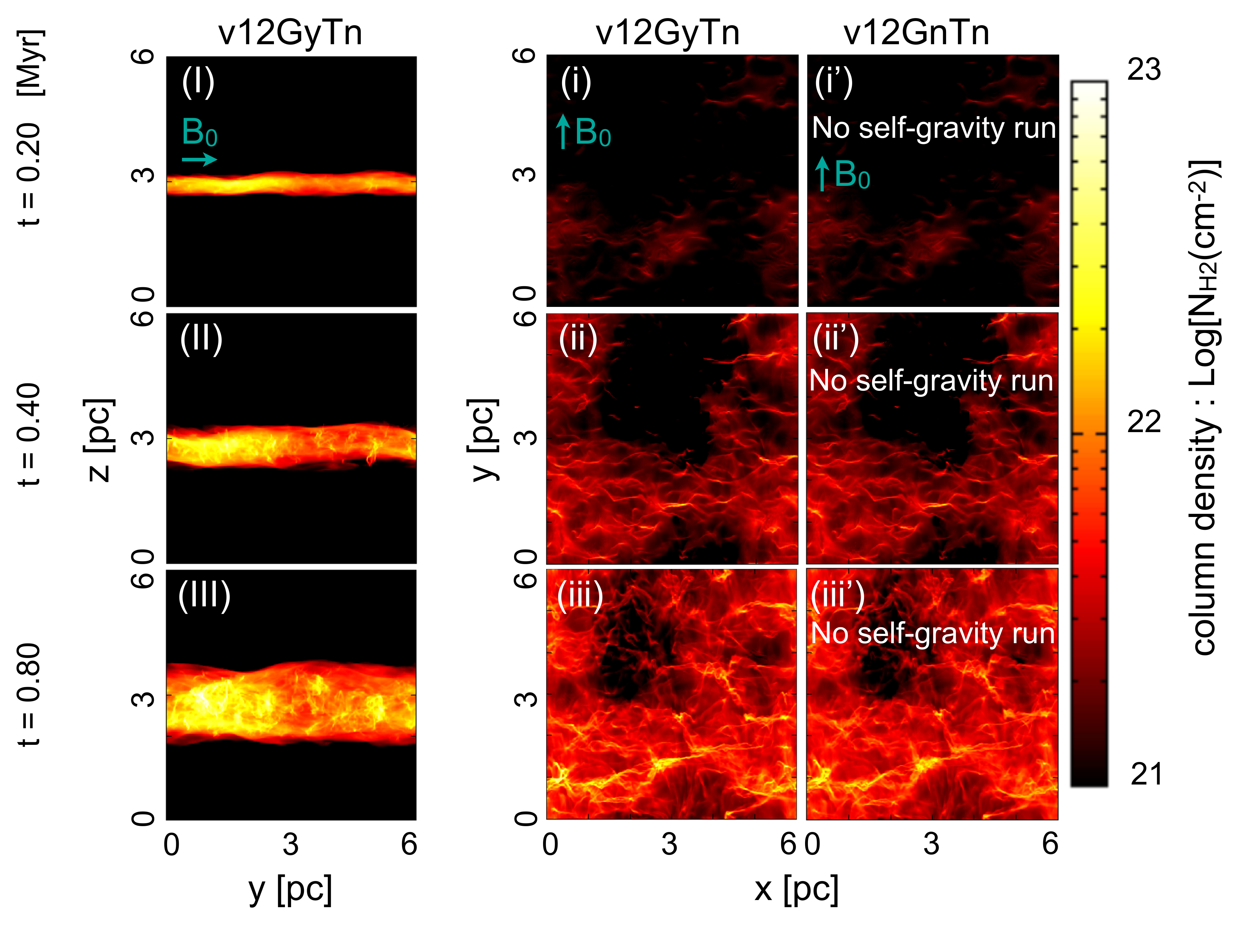}
    \caption{\small{Column density maps at time $t = 0.2\ (top),\ 0.4\ (middle),\ \mathrm{and}\ 0.8\ (bottom)\ \mathrm{Myr}$.
    \textit{Left row }(panels I, II, and III): Column density in the $y$-$z$ plane of model v12GyTn.
    \textit{Middle row }(panels i, ii, and iii): Same as panels (I)-(III) but for the $x$-$y$ plane.
    \textit{Right row }(panels i', ii', and iii'): Same as panels (i)-(iii) but for  model v12GnTn.
    }}
    \label{fig:v12n1}
\end{center}
\end{figure*}
\begin{figure*}
    \begin{center}
        \includegraphics[clip,width=17.5cm]{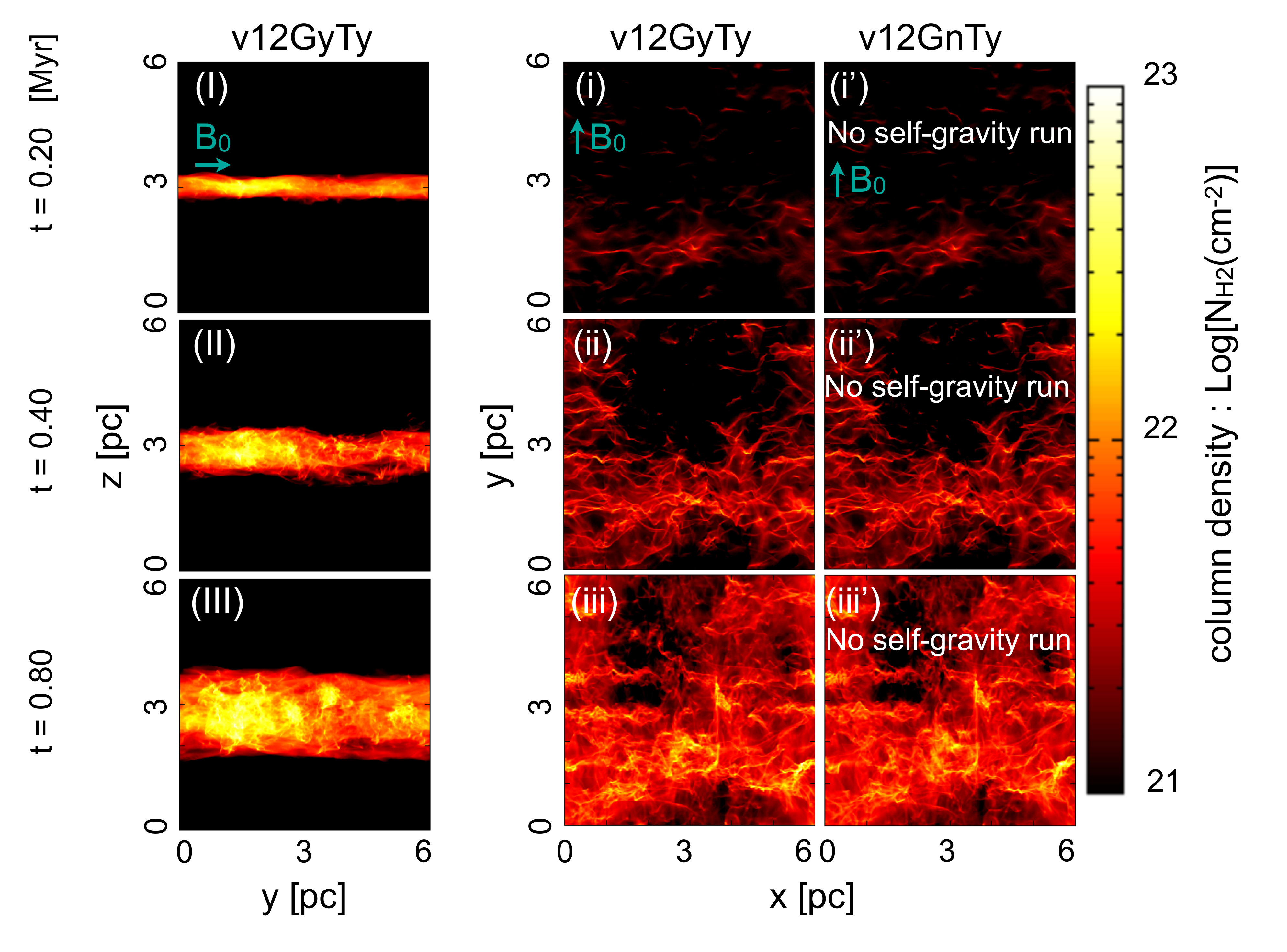}
        \caption{\small{Column density maps at time $t = 0.2\ (top),\ 0.4\ (middle),\ \mathrm{and}\ 0.8\ (bottom)\ \mathrm{Myr}$.
        \textit{Left row }(panels I, II, and III): Column density in the $y$-$z$ plane of model v12GyTy.
        \textit{Middle row }(panels i, ii, and iii): Same as panels (I)-(III) but for the $x$-$y$ plane.
        \textit{Right row }(panels i', ii', and iii'): Same as panels (i)-(iii) but for model v12GnTy.
        }}
        \label{fig:v12n1+tb}
    \end{center}
\end{figure*}
\begin{figure*}
    \begin{center}
        \includegraphics[clip,width=18cm]{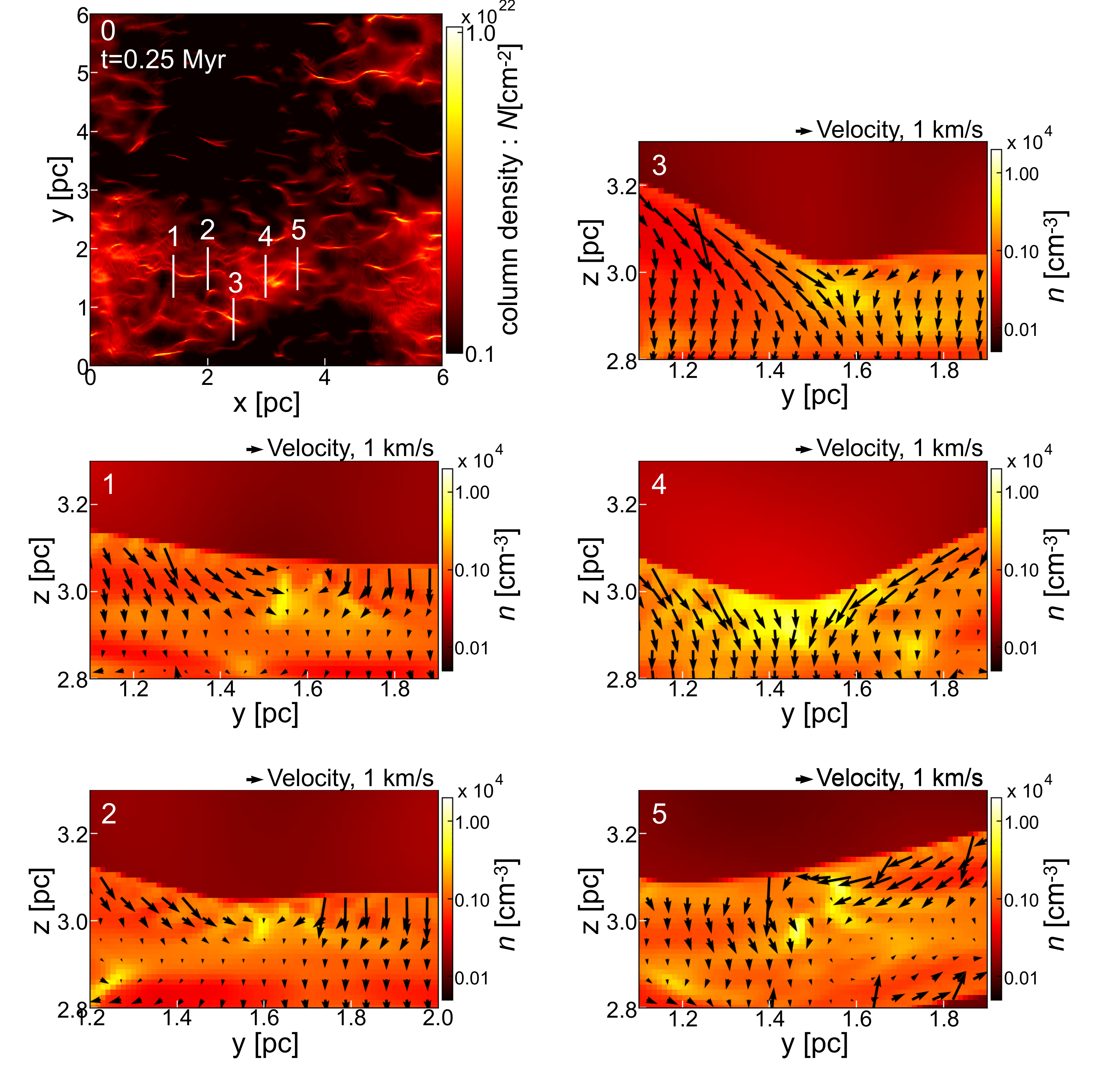}
        \caption{\small{
        Panel (0): Early stage ($t=0.25\ \mathrm{Myr}$) column density map in the $x$-$y$ plane of result of model v12GyTn.
        The five white lines mark the planes in which the cross-section maps in panels (1)-(5) are drawn.
        Panels (1)-(5): Cross-section maps of the number density in the $y$-$z$ plane.
        The yellow blobs located roughly at the center of each panel correspond to cross-sections of the filaments formed by type O mechanism.
        }}
        \label{fig:pvv12Tn}
    \end{center}
\end{figure*}
\begin{figure*}
    \begin{center}
        \includegraphics[clip,width=18cm]{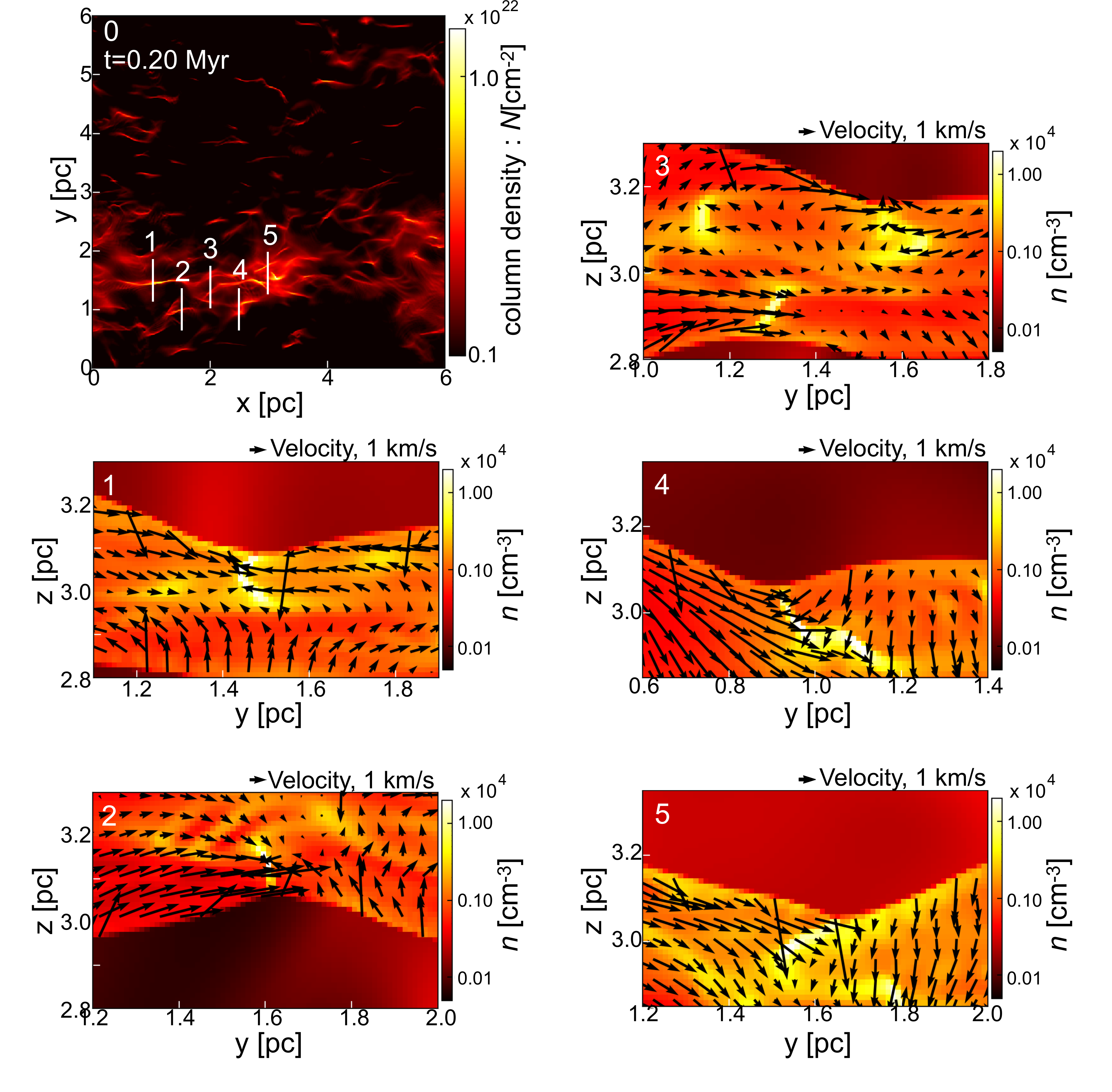}
        \caption{\small{
        Same as Figure \ref{fig:pvv12Tn} but for model v12GyTy. We can confirm that type O mechanism takes a major role in the filament formation even in the case with initial turbulence.
        }}
        \label{fig:pvv12Ty}
    \end{center}
\end{figure*}
\begin{figure*}
    \begin{center}
        \includegraphics[clip,width=17.5cm]{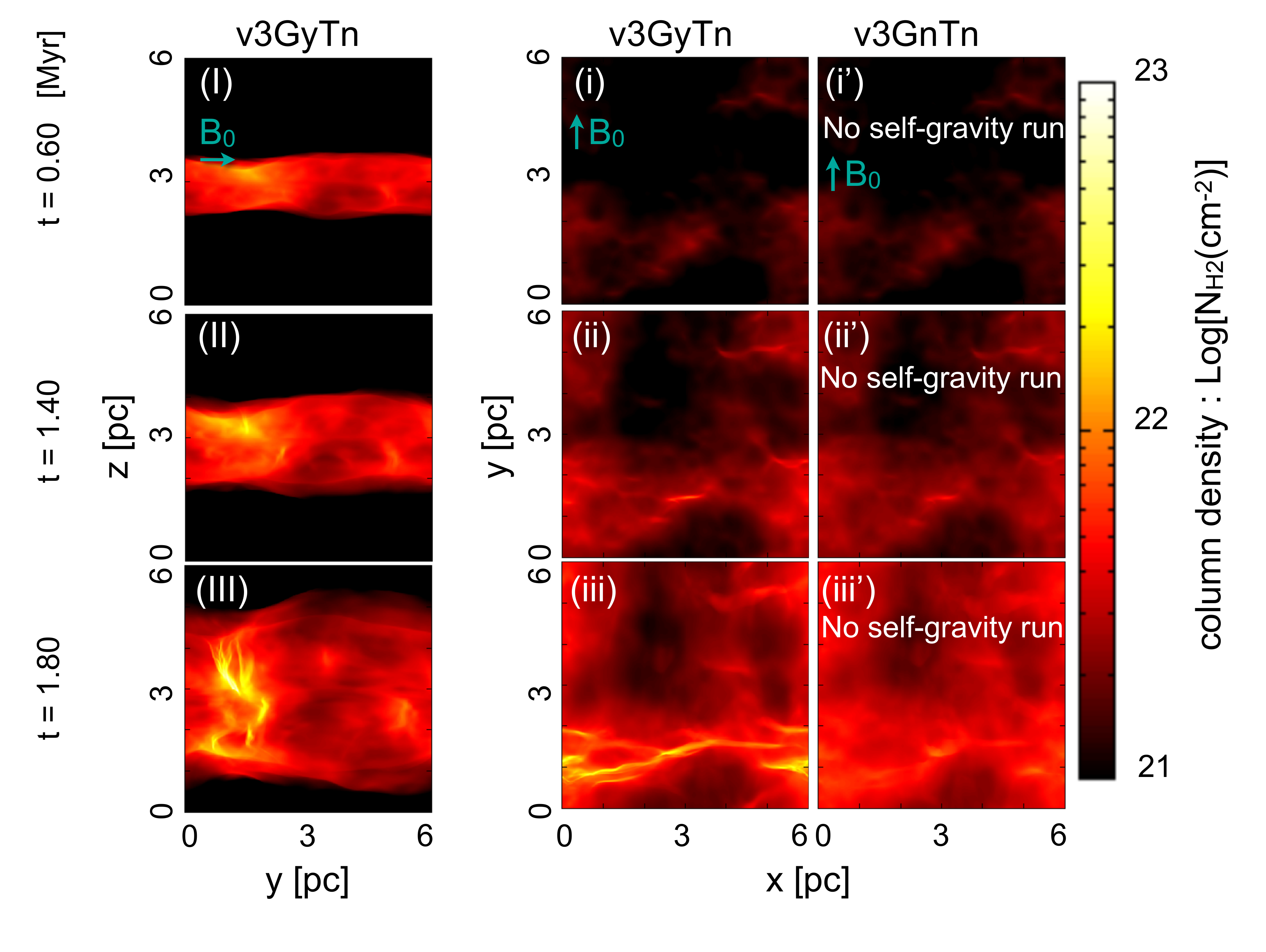}
        \caption{\small{Column density maps at time $t = 0.6\ (top),\ 1.4\ (middle),\ \mathrm{and}\ 1.8\ (bottom)\ \mathrm{Myr}$.
        \textit{Left row }(panels I, II, and III): Column density in the $y$-$z$ plane of model v3GyTn.
        \textit{Middle row }(panels i, ii, and iii): Same as the panels (I)-(III) but for the $x$-$y$ plane.
        \textit{Right row }(panels i', ii' , and iii'): Same as the panels (i)-(iii) but for model v3GnTn.
        }}
        \label{fig:v3n1}
    \end{center}
\end{figure*}
\begin{figure*}
    \begin{center}
        \includegraphics[clip,width=17.5cm]{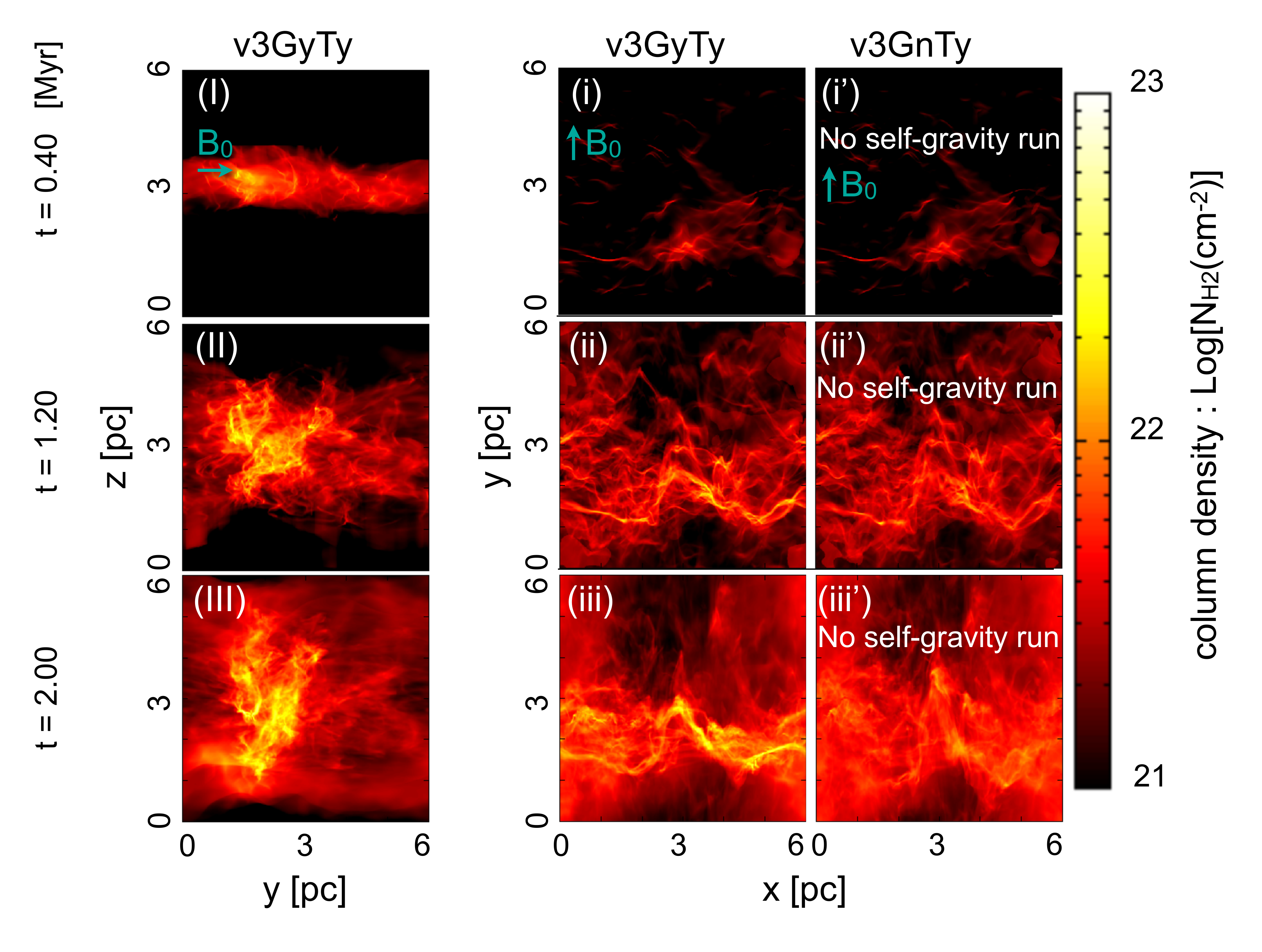}
        \caption{\small{Column density maps at time $t = 0.4\ (top),\ 1.2\ (middle),\ \mathrm{and}\ 2.0\ (bottom)\ \mathrm{Myr}$.
        \textit{Left row }(panels I, II, and III): Column density in the $y$-$z$ plane of model v3GyTy.
        \textit{Middle row }(panels i, ii, and iii): Same as panels (I)-(III) but for the $x$-$y$ plane.
        \textit{Right row }(panels i', ii', and iii'): Same as panels (i)-(iii) but for model v3GnTy.
        }}
        \label{fig:v3n1+tb}
    \end{center}
\end{figure*}
\begin{figure}
    \begin{center}
        \includegraphics[clip,width=8.0cm]{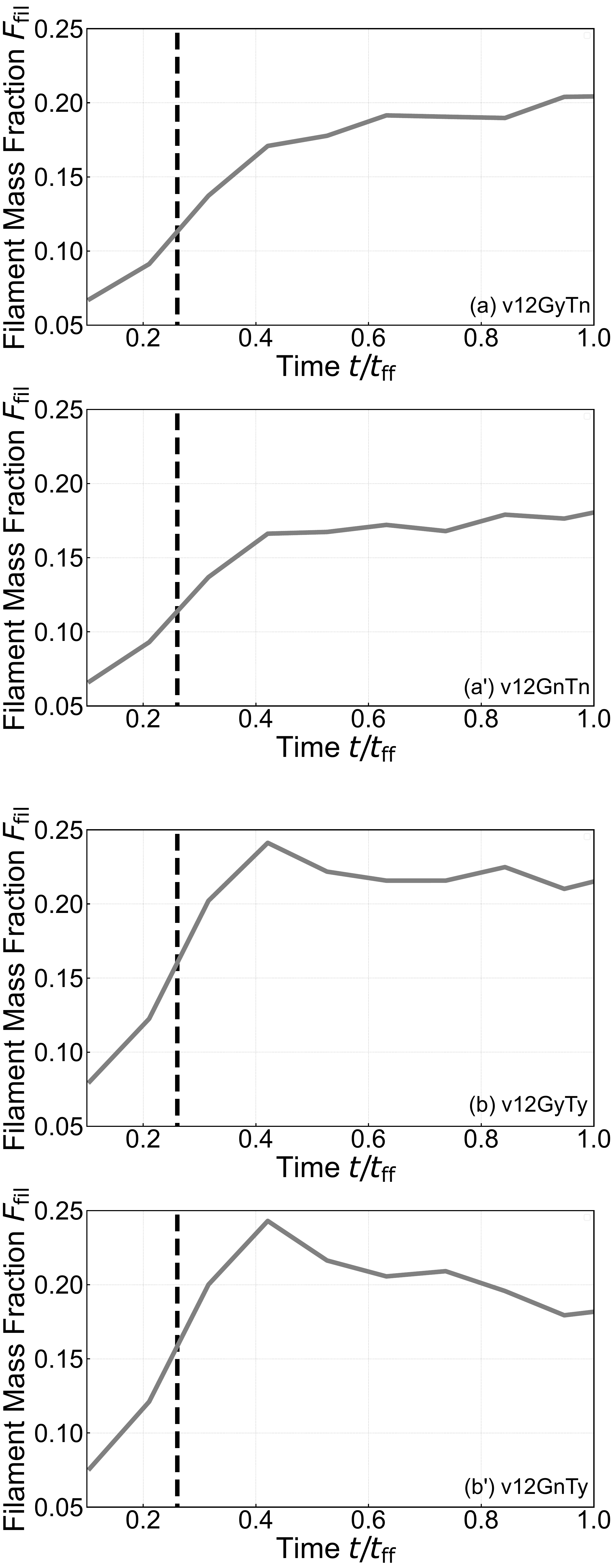}
        \caption{\small{Temporal evolution of the filament mass fraction $F_{\mathrm{fil}}$ (solid line) and the filament formation time $t_{\mathrm{fil}}$ (dashed line).
        Panel (a), (a'), (b), and (b') are the results of models v12GyTn, v12GnTn, v12GyTy, and v12GnTy, respectively.
        }}
        \label{fig:tfilv12}
    \end{center}
\end{figure}
\begin{figure}
    \begin{center}
        \includegraphics[clip,width=8.0cm]{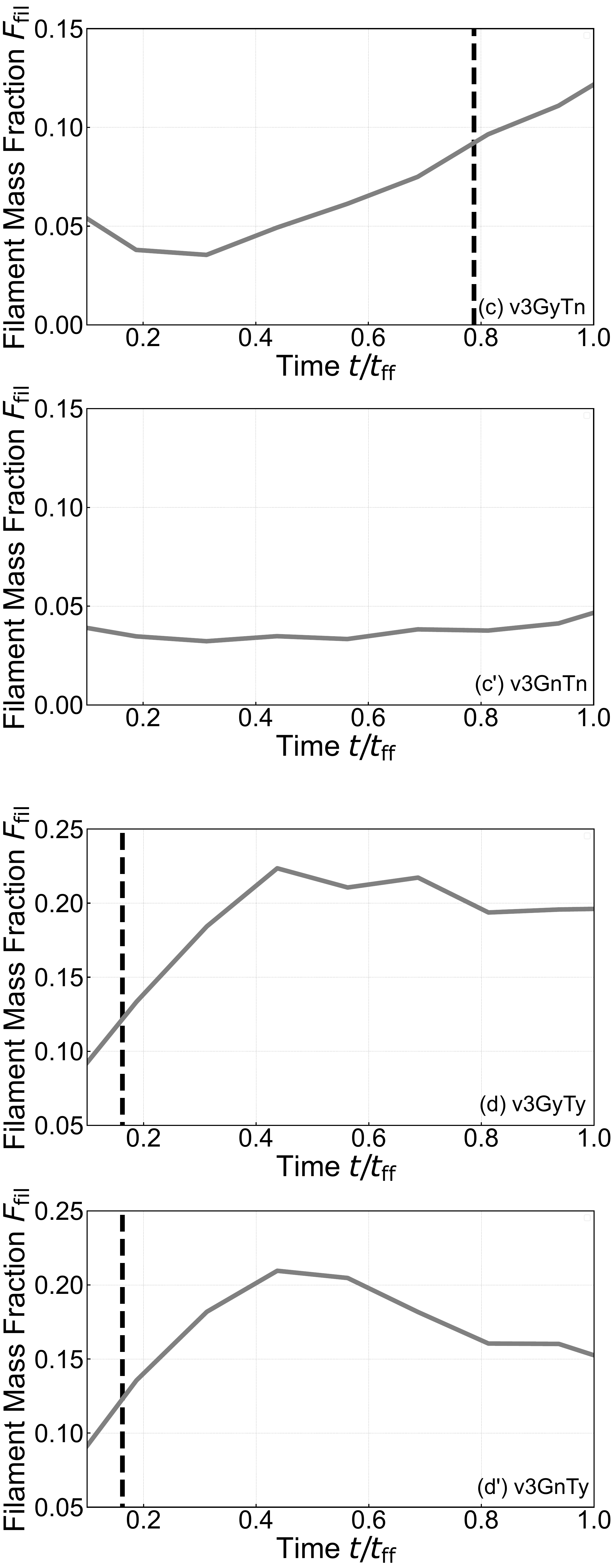}
        \caption{\small{Temporal evolution of the filament mass fraction $F_{\mathrm{fil}}$ (solid line) and the filament formation time $t_{\mathrm{fil}}$ (dashed line).
        Panels (c), (c'), (d), and (d') are the results of models v3GyTn, v3GnTn, v3GyTy, and v3GnTy, respectively.
        }}
        \label{fig:tfilv3}
    \end{center}
\end{figure}
\begin{figure}
    \begin{center}
        \includegraphics[clip,width=8.0cm]{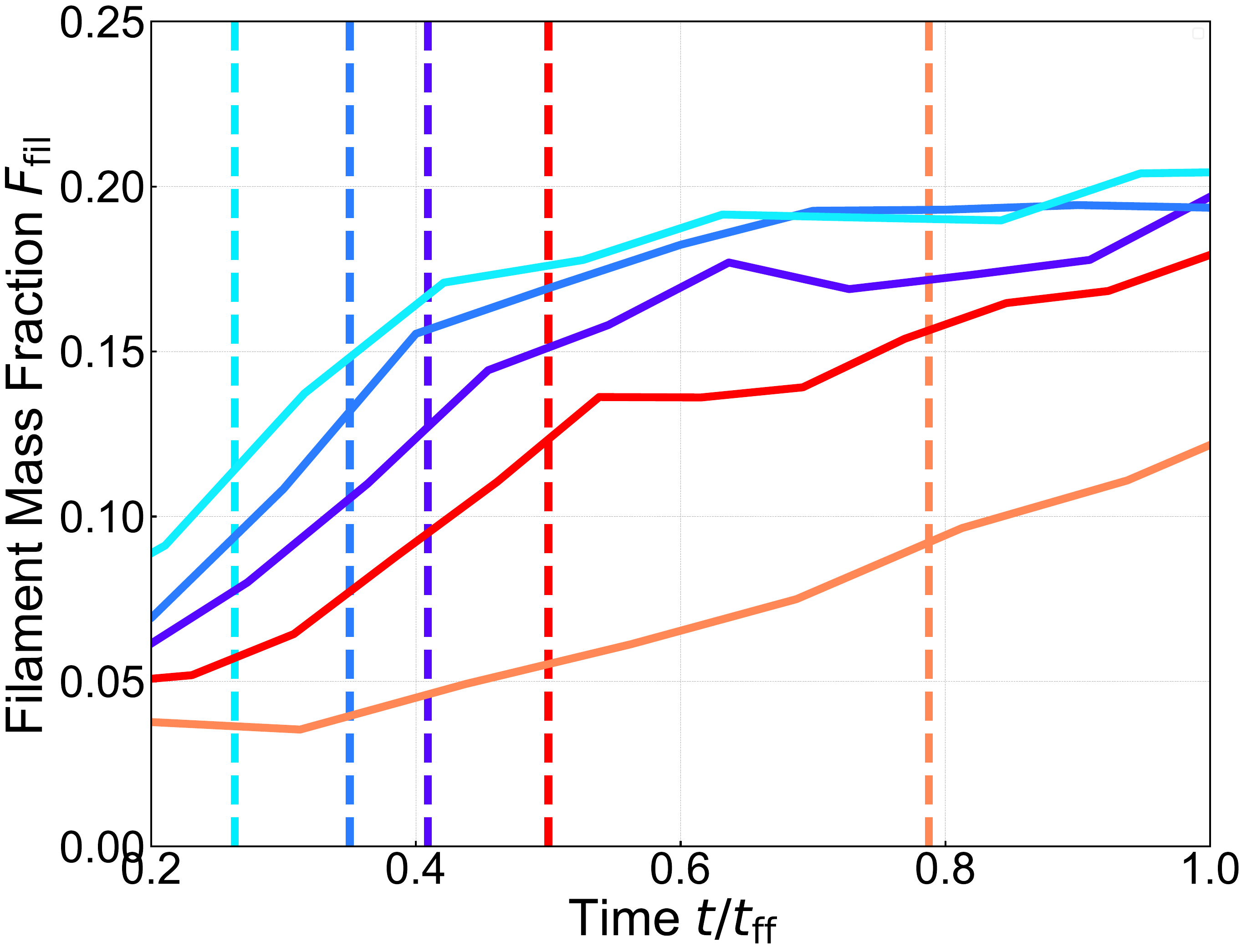}
        \caption{\small{Temporal evolution of the filament mass fraction $F_{\mathrm{fil}}$ (solid lines) and the filament formation time $t_{\mathrm{fil}}$ (dashed lines) in models with self-gravity and various shock velocities but without turbulent velocity fluctuation. 
        The colors show the results for models v12GyTn (cyan), v10GyTn (blue), v8GyTn (purple), v6GyTn (red), and v3GyTn (orange).
        }}
        \label{fig:fraction_th}
    \end{center}
\end{figure}

\begin{figure*}
    \begin{center}
        \includegraphics[clip,width=18.05 cm]{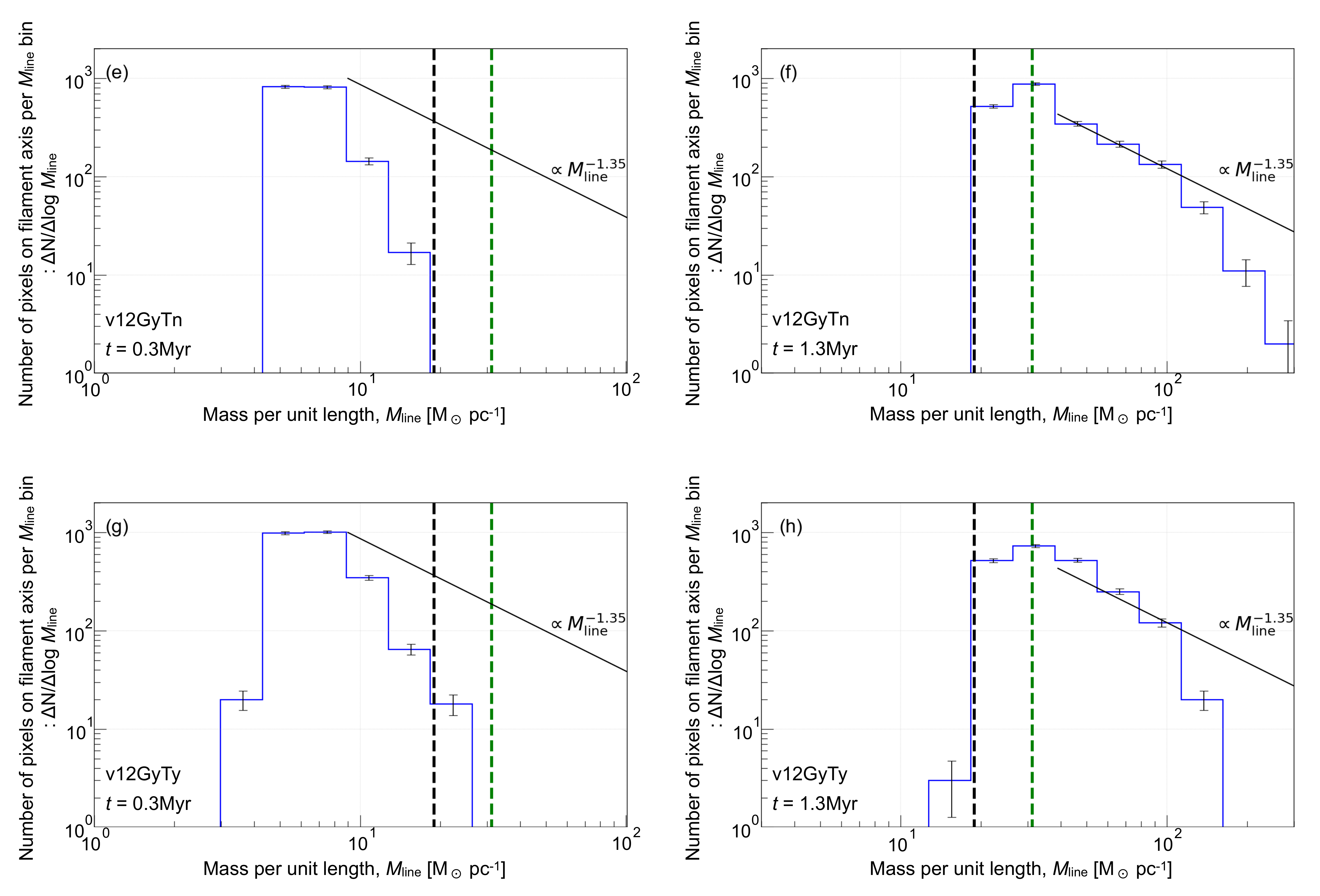}
        \caption{\small{Filament mass functions for model v12GyTn (\textit{top panels} (e) and (f)) and v12GyTy (\textit{bottom panels} (g) and (h)).
        \textit{Left panels} (e) and (g): Filaments mass functions at time $t=t_{\mathrm{fil}}$.
        \textit{Right panels} (f) and (h): Results at time $t>\ 1\ \mathrm{Myr}$.
        The black solid line represents the power-law function of the Salpeter initial mass function; the black dashed line shows the critical line-mass~\protect\citep{Stodolkiewicz1963,Ostriker1964}; and the green dashed line is the critical line-mass considering the magnetic field~\protect\citep{Tomisaka2014}.
        }}
        \label{fig:lmfv12}
    \end{center}
\end{figure*}
\begin{figure*}
    \begin{center}
        \includegraphics[clip,
        width=18.05 cm]{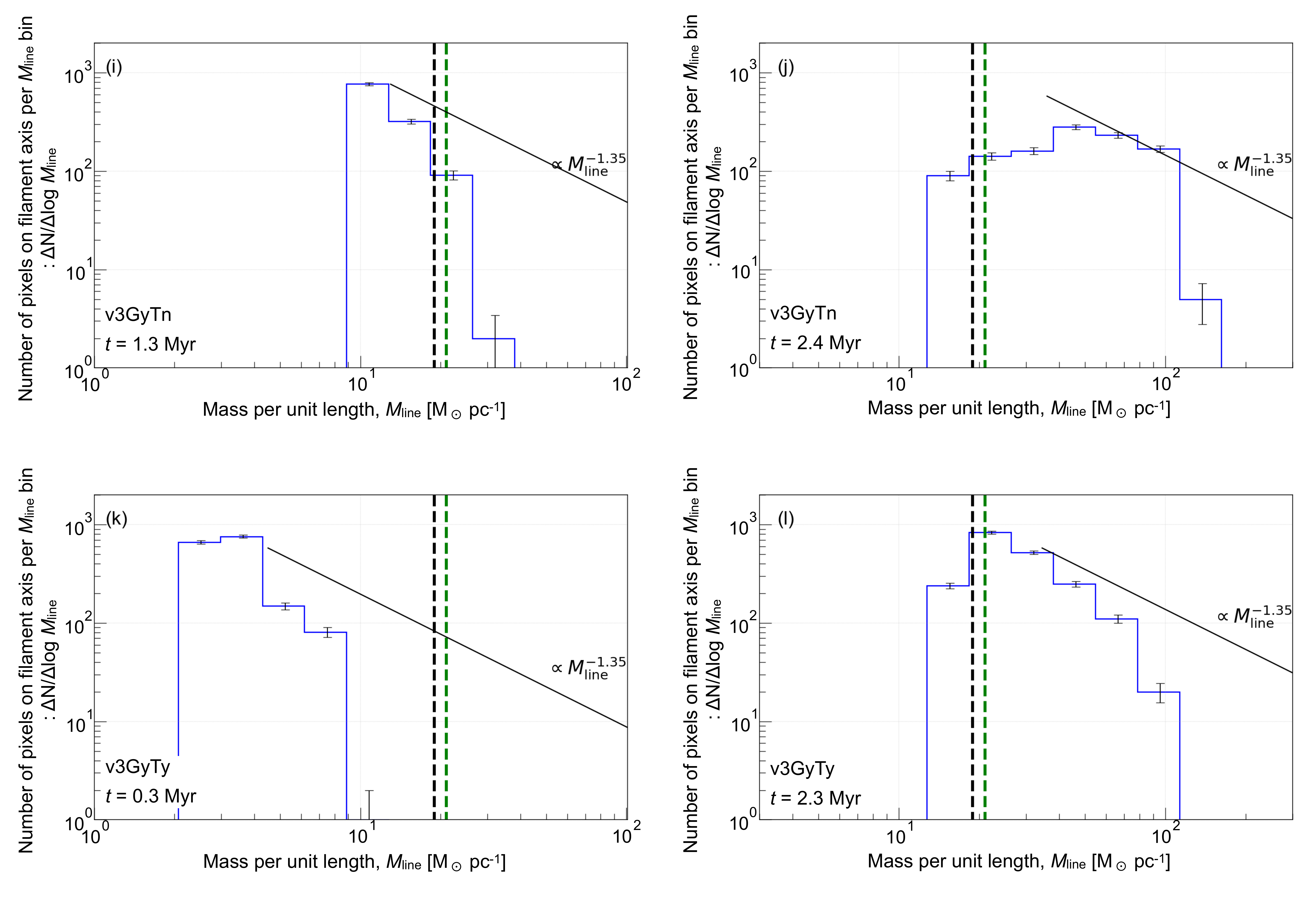}
        \caption{\small{Filament mass functions for model v3GyTn (\textit{top panels} (i) and (j)) and v3GyTy (\textit{bottom panels} (k) and (l)).
        \textit{Left panels} (i) and (k): Filament mass functions at time $t=t_{\mathrm{fil}}$.
        \textit{Right panels} (j) and (l): Results at time $t>\ 2\ \mathrm{Myr}$.
        The lines colors are the same as those defined in Figure \ref{fig:lmfv12}.
        }}
        \label{fig:lmfv3}
    \end{center}
\end{figure*}
\begin{figure*}
    \begin{center}
        \includegraphics[clip,
        width=16.05 cm]{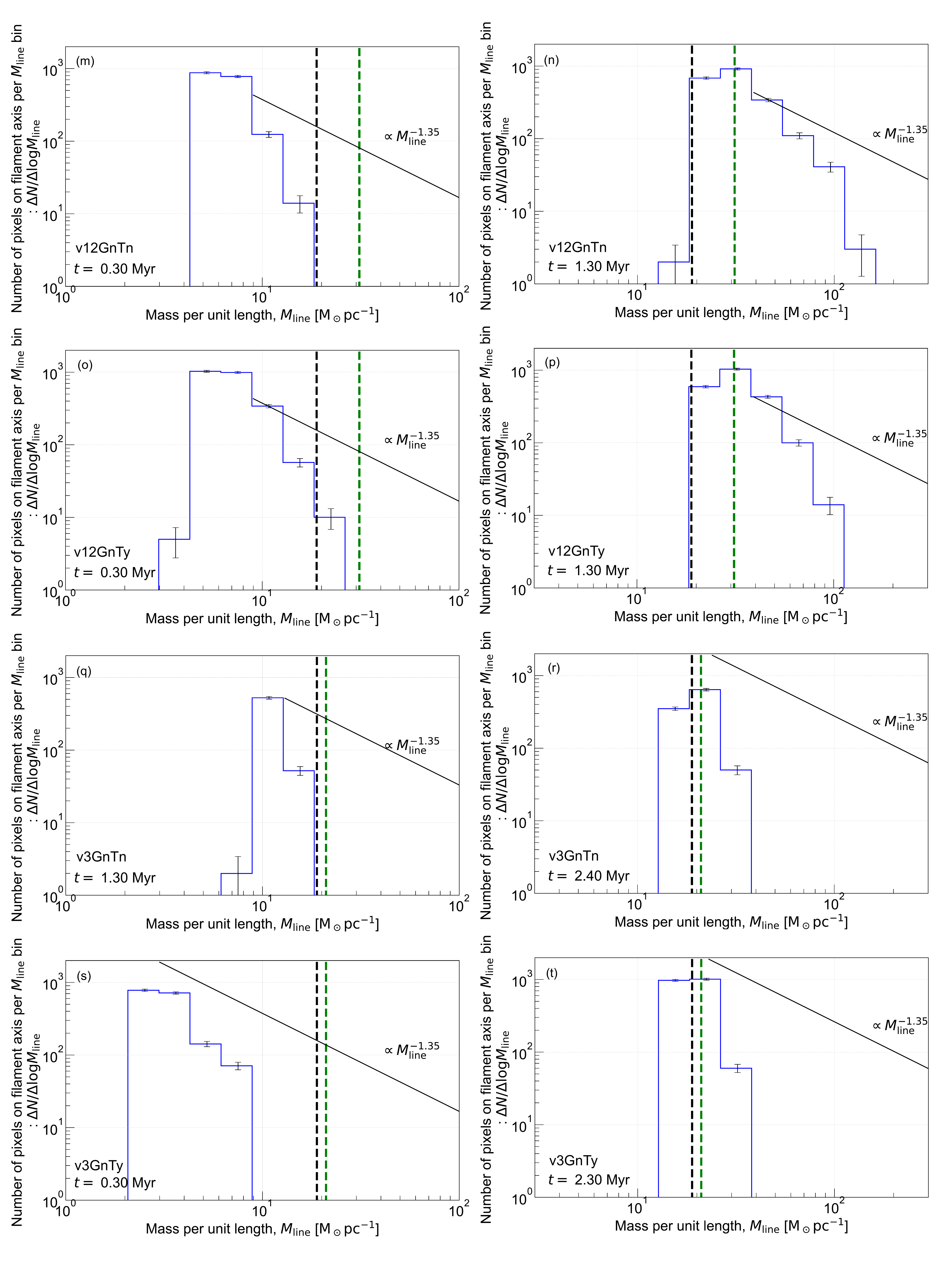}
        \caption{\small{Filament mass functions for no-self-gravitating models v12GnTn (panels m and n), v12GnTy (panels o and p), v12GnTn (panels q and r), and v12GnTy (panels s and t).
        To compare the histograms in the models with and without self-gravity, we show the histograms that are taken at the same times in Figure \ref{fig:lmfv12} and \ref{fig:lmfv3}.
        The lines colors are the same as those defined in Figure \ref{fig:lmfv12}.
        }}
        \label{fig:lmf_nog}
    \end{center}
\end{figure*}
\begin{figure*}
    \begin{center}
        \includegraphics[clip,
        width=18.05 cm]{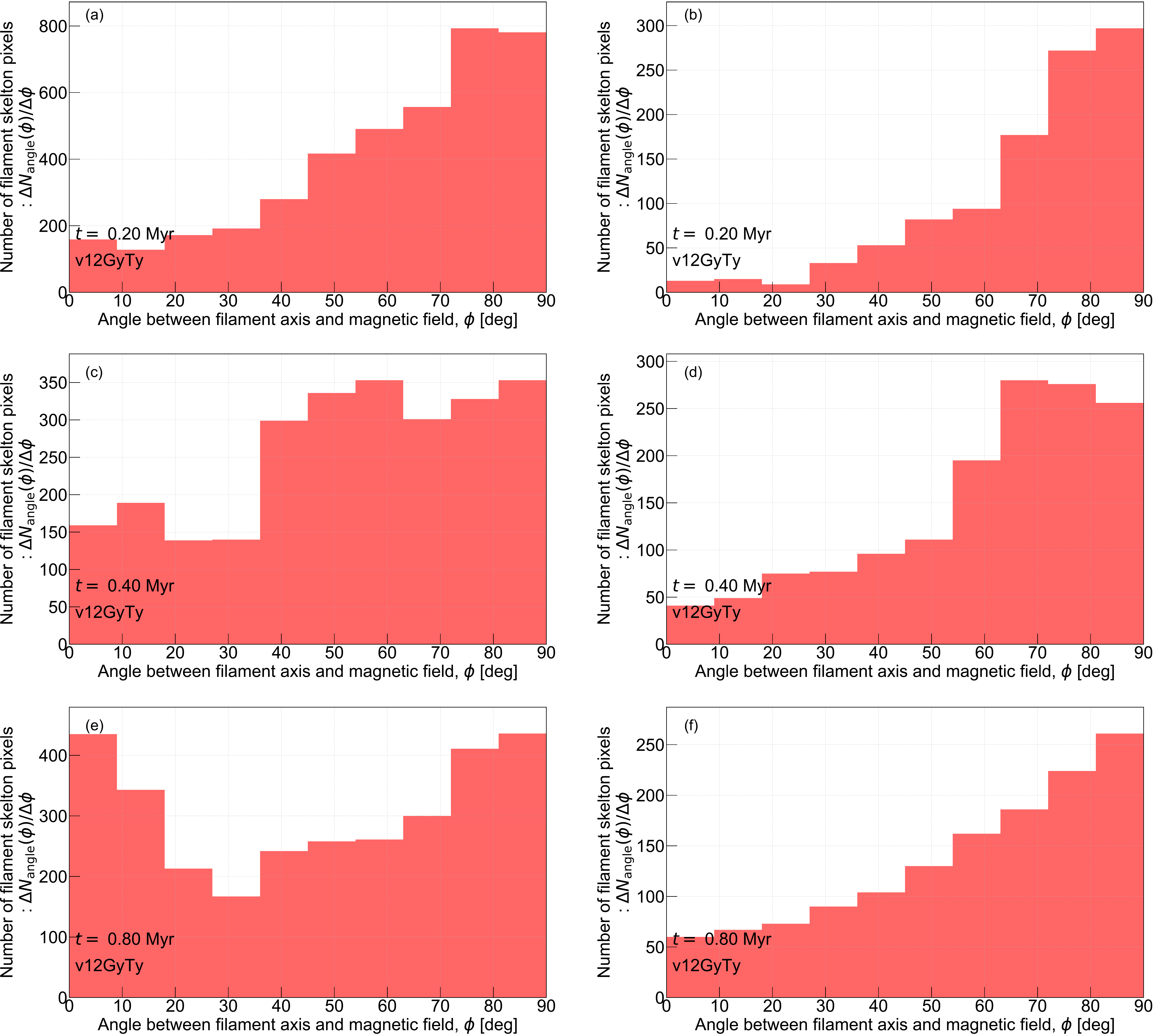}
        \caption{\small{Histogram of angles between filaments and magnetic field for model v12GyTy.
        From top to bottom, results at time $t$=\ 0.2, 0.4 and 0.8 $\mathrm{Myr}$, respectively. \textit{Top panels} (1) and (2) are results at filament formation time (see \S \ref{subsec:Filament Formation Timescale} and Figure \ref{fig:tfilv12}).
        \textit{Bottom panels} (5) and (6) are results at close to free-fall time~(Eq. \ref{equation:freefalltime}).
        \textit{Left panels} (1), (3), and (5): Results when we identify filaments in the column density range of 0.5$\bar{N}_{\mathrm{sh}}$ to 1.5$\bar{N}_{\mathrm{sh}}$.
        \textit{Right panels} (2), (4), and (6): Results when the filament identification threshold column density is chosen to be 1.5$\bar{N}_{\mathrm{sh}}$.
        }}
        \label{fig:angle_v12}
    \end{center}
\end{figure*}
\begin{figure*}
    \begin{center}
        \includegraphics[clip,
        width=18.05 cm]{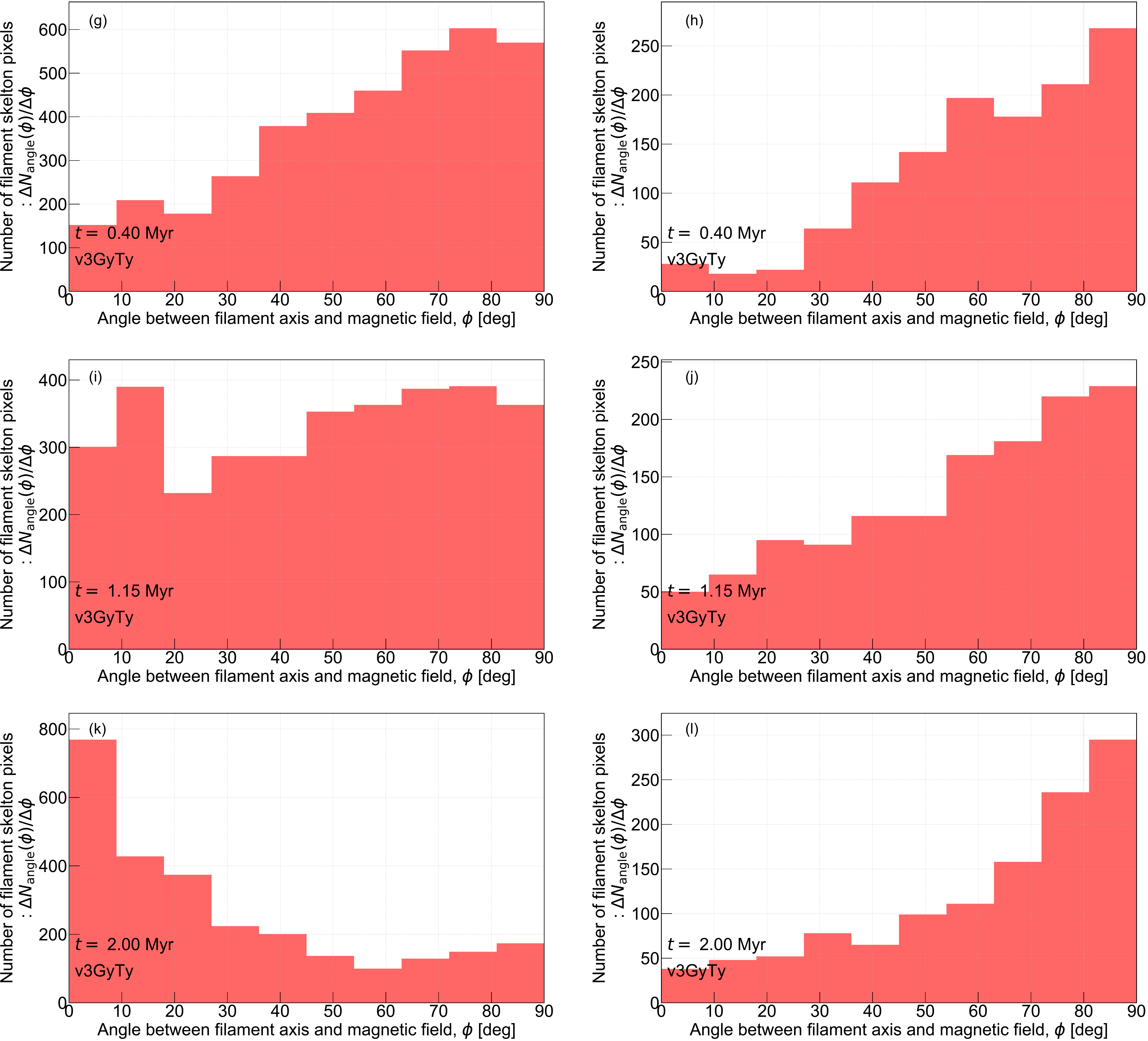}
        \caption{\small{Histogram of angles between filaments and magnetic field for model v3GyTy.
        From top to bottom, results at time $t$=\ 0.4, 1.15 and 2.0 $\mathrm{Myr}$, respectively.
        \textit{Top panels} (7) and (8) are results at filament formation time (see \S \ref{subsec:Filament Formation Timescale} and Figure \ref{fig:tfilv3}).
        \textit{Bottom panels} (11) and (12) are results which exceed free-fall time~(Eq. \ref{equation:freefalltime}).
        \textit{Left panels} (7), (9), and (11): Results when we identify filaments in the column density range of 0.5$\bar{N}_{\mathrm{sh}}$ to 1.5$\bar{N}_{\mathrm{sh}}$.
        \textit{Right panels} (8), (10), and (12): Results when the filament identification threshold column density is chosen to be 1.5$\bar{N}_{\mathrm{sh}}$.
        }}
        \label{fig:angle_v3}
    \end{center}
\end{figure*}

To investigate its effect on filament formation, we also perform simulations with and without self-gravity.
When we take into account the self-gravity, we use the sink particle technique in the regions in which gravitational collapse is expected to occur.
The sink particle generation condition is the same as that discussed in previous research~\protect\citep[][]{Inoue2018,Matsumoto2015}.
It should be noted that the employment of the sink particle is simply for a continuation of the simulations even after onset of local gravitational collapse\footnote{The threshold density for the sink formation is $5.6\times10^4$ cm$^{-3}$. This value is much lower than that used in our previous studies, because we do not employ the adaptive mesh refinement in the present study}.
In this paper, we will not focus on the information of sink particles, because resulting total mass of the sink particles are much smaller than that of dense filaments (for instance, the final total mass of the sink particles is only 3\% to that of dense filaments for model v12GyTy, and 6\% for model v3GyTy).

\section{Results} \label{sec:Results}
\subsection{High Shock Velocity Case} \label{subsec:fast coll case}

In the case of $v_{\mathrm{coll}} = 12\ \mathrm{km\ s^{-1}}$, the flow collision induces fast isothermal MHD shock waves.
According to the shock jump condition \protect\citep[e.g., \S 4.1 of ][]{Fukui2020}, the compression ratio of the isothermal fast shock is given by
\begin{equation}
    \rho_1 / \rho_0 = v_0 / v_1 \simeq \sqrt{2} \mathcal{M}_{\mathrm{A}},
    \label{equation:shock jump cond}
\end{equation}
where $\mathcal{M}_A \equiv v_0/v_{\mathrm{Alf},0}$ is the Alfv\'enic Mach number, the Alfv\'en velocity is given by $v_{\mathrm{Alf},0}=B_0/\sqrt{4\pi \rho_0}$ and subscripts 0 and 1 indicate preshock and postshock values respectively.
The preshock velocity in the shock rest-frame $v_0$ is equivalent to the shock velocity in the upstream rest frame $v_{\mathrm{sh}}$.
Given that the numerical domain is on the postshock rest frame, the shock wave propagates with the velocity of $v_1 \simeq v_0 / (\sqrt{2} \mathcal{M}_{\mathrm{A}})$.
The relation between the converging flow velocity $v_{\mathrm{coll}}$ and the shock velocity $v_{\mathrm{sh}}$ is given by
\begin{eqnarray}
    v_{\mathrm{sh}} &=& v_{\mathrm{coll}}/2 + v_1 \nonumber \\ 
    &\simeq& v_{\mathrm{coll}}/2 + v_{\mathrm{Alf,0}}/\sqrt{2} \nonumber \\ 
    &\simeq& 6 \left( \frac{v_{\mathrm{coll}}}{12\ \mathrm{km\ s^{-1}}} \right)\ \mathrm{km\ s^{-1}} \nonumber \\ 
    & &+ 1 \left( \frac{B_0}{10\ \mathrm{\mu G}} \right) \left( \frac{n_0}{100\ \mathrm{cm^{-3}}} \right)^{-1/2}\ \mathrm{km\ s^{-1}}.
    \label{equation:shock velocity}
\end{eqnarray}
In Figure \ref{fig:v12n1}, we show snapshots of the column density structure of the results of models v12GyTn and v12GnTn at $t$ = 0.2 (top), 0.40 (middle), and 0.60 (bottom) Myr.
In constructing the column density, we identify the shock fronts and integrate the density only in the shock-compressed region.
To identify the shock fronts, we scan the total pressure  $\rho (x,y,z) c_{\mathrm{s}}^{2}+B^2(x,y,z)/8\pi$ along the z-axis from upstream ($z=0$ and $6$) to downstream ($z=3$) and define the two shock fronts as the largest and smallest locations of $z(x,y)$ where the following condition is satisfied:
\begin{equation}
    \rho (x,y,z) c_{\mathrm{s}}^{2}+B^{2}(x,y,z)/8\pi \geq f_{\mathrm{jump}} \times \bar{\rho}_0 v_{\mathrm{coll}}^{2}.
    \label{equation:shock id cond}
\end{equation}
Here, we choose $f_{\mathrm{jump}}$ = 0.2 for convenience, but we confirmed that the result with $f_{\mathrm{jump}}$=0.4 gives the indistinguishable result to $f_{\mathrm{jump}}$ = 0.2 case.

Panels (I)-(III) and (i)-(iii) in Figure \ref{fig:v12n1} show column density snapshots of model v12GyTn in the $y$-$z$ and $x$-$y$ planes, respectively.
Panels (I), (II), and (III) show two shock waves induced by the converging flows propagating toward the positive and negative $z$-directions.
Panels (i)-(iii) and (i')-(iii') show the formation of many filaments regardless of the effect of self-gravity that indicates the filament formation is not driven by self-gravity in the present high shock velocity cases.
In \S \ref{subsec:Filament Line Mass Function}, we show that very massive filaments as large as $100\ \mathrm{M_{\odot}\ pc^{-1}}$ are formed at $t$ = 1 Myr in this series of high shock velocity results.

Because the present models (v12GyTn and v12GnTn) do not include initial turbulence, type O mechanism definitively accounts for the filament formation.

In Figure \ref{fig:v12n1+tb}, we show snapshots at $t$ = 0.20 (top), 0.40 (middle), and 0.80 (bottom) Myr of models v12GyTy (panels I-III and i-iii) and v12GnTy (panels i'-iii').
In panels (i)-(iii) and (i')-(iii'), more (mostly faint) filaments parallel to the mean magnetic field lines are present compared with that shown in Figure \ref{fig:v12n1}.
This indicates that type C mechanism helps to create (faint) filaments.
By comparing these two models, we again find that type G does not account for the filament formation.
However, see below for Figure \ref{fig:tfilv12} where we will see that the resulting filaments tend to disperse in the absence of self-gravity.  

To clarify the dominant filament formation mechanism, we show the local density cross-sections around the five major filaments as the results of models v12GyTn in Figure \ref{fig:pvv12Tn} and v12GyTy in Figure \ref{fig:pvv12Ty}.
The high density blobs (yellow regions) in the local cross-section maps correspond to those of the major filaments.
In type O mechanism, the post-shock gas flows toward a region behind the concave shock surface, where the filament is formed.
Such a characteristic flow is created due to the effect of oblique MHD shock wave.
Further details of the flow structure have been reported by \protect\citet{Inoue2013} and \protect\citet{Inoue2018}.
The curved shock morphology and velocity vectors (black arrows) shown in the cross-section panels in both v12GyTn and v12GyTy models clearly support type O origin of the (major) filaments.

In principle, the flows in the post-shock region induced by the oblique MHD shock may drive turbulence in the long-term evolution.
However, at least in the stage of filament formation, the converging flows along the bending magnetic field is laminar.
Thus, it is not appropriate to simply say that type O filament formation is involved in the simulation of supersonic turbulence in general.
Type O mechanism expected to selectively appears in the compression by a shock waves with a relatively smooth surface, such as the one caused by an expansion of the super-shell or HII region.
It should be noted that we can observationally distinguish type O mechanism from the other types by measuring the structure of magnetic and velocity fields~\protect\citep{Arzoumanian2018,Tahani2018,Tahani2019,Bonne2020,Chen2020,Kandori2020a,Kandori2020b}.

\subsection{Low Shock Velocity Case} \label{subsec:slow coll case}
In Figure \ref{fig:v3n1}, we show snapshots at $t$ = 0.6 $\mathrm{(top)}$, 1.4 $\mathrm{(middle)}$, and 1.8 $\mathrm{(bottom)}$ Myr of models v3GyTn (panels I-III and i-iii) and v3GnTn (panels i'-iii').
From Eq. (\ref{equation:shock velocity}), the average shock velocity is calculated to be $2.5\ \mathrm{km\ s^{-1}}$ for $v_{\mathrm{coll}} = 3\ \mathrm{km\ s^{-1}}$ simulations presented in this section.
Although filamentary structures are created in panel (iii), no obvious dense filaments are shown in panel (iii').
This suggests that type O mechanism does not work for this low shock velocity case and that self-gravity (type G) accounts for the filament formation in model v3GyTn.

Figure \ref{fig:v3n1+tb} shows snapshots at $t$ = 0.4 $\mathrm{(top)}$, 1.20 $\mathrm{(middle)}$, and 2.00 $\mathrm{(bottom)}$ Myr of model v3GyTy (panels I-III and i- iii) and the model v3GnTy (panels i'-iii').
The similar filamentary structure formation occurring at $t$ = 0.4 and 1.2 Myr in both models suggests that type C filament formation is important for the low shock velocity models\footnote{Even in the low shock velocity models, some major filaments appear to be formed by type O mechanism.}.
In the later stage of $t$ = 2.0 Myr, the results of model v3GyTy show that the filaments are attracted to each other by the self-gravity that eventually induces filament collisions and enhances the filament line-mass.
More detailed analysis is given in \S \ref{subsec:Filament Line Mass Function}.

\subsection{Filament Formation Timescale vs. Free-Fall Time} \label{subsec:Filament Formation Timescale}
In this section, we use the following procedure to compute the filament formation time.
First, we identify the filaments by employing the FilFinder algorithm~\protect\citep{Koch2015} that returns filament skeletons for a given input two-dimensional image.
The skeleton is a single-pixel-width object that corresponds to the major-axis of a filament.
To focus on major filaments, we neglect filaments having column density smaller than $1.5\times \bar{N}_{\mathrm{sh}}$, where $\bar{N}_{\mathrm{sh}}$ is the mean column density of the shocked region.
We confirmed that the result does not change even if we change the factor 1.5 to 2.0.
We stress that because of this minimum column density requirement for the filament identification, our analysis given below always omits faint filaments with column densities smaller than $1.5\times \bar{N}_{\mathrm{sh}}$.

Then, we calculate the filament mass fraction $F_{\mathrm{fil}}$ defined by
\begin{equation}
    F_{\mathrm{fil}} \equiv M_{\mathrm{fil,tot}}/M_{\mathrm{sh}},
\end{equation}
where $M_{\mathrm{fil,tot}}$ and $M_{\mathrm{sh}}$ are the total mass of the filaments in the snapshot and the shocked region mass, respectively.
$M_{\mathrm{fil,tot}}$ is computed by integrating the gas column density over the region around 0.1 pc of the filament skeleton~\protect\citep{Arzoumanian2011,Koch2015}.
Using the $F_{\mathrm{fil}}$, we define the filament formation time $t_{\mathrm{fil}}$ as the time at which the filaments are produced most actively.
Specifically, The time at which $\Delta F_{\mathrm{fil}}/\Delta t$ reaches its maximum value is defined as the $t_{\mathrm{fil}}$.
Where $\Delta t$ and $\Delta F_{\mathrm{fil}}$ are the one-tenth of free-fall time $t_{\mathrm{ff}}$ in the shocked layer and the increment of $F_{\mathrm{fil}}$ in the time interval $\Delta t$, respectively.

The free-fall time in the post-shock layer, which gives the timescale of the self-gravitating sheet fragmentation~\protect\citep{Nagai1998}, can be estimated as
\begin{eqnarray}
    t_{\mathrm{ff}}
&=&
\sqrt{\frac{1}{2\pi G\bar{\rho}_1}}\nonumber \\
&=&
\sqrt{\frac{\bar{v}_{\mathrm{Alf}}}{2 \sqrt{2}\pi G \bar{\rho}_0 \bar{v}_{\mathrm{sh}}}}\nonumber \\
&=&
\sqrt{\frac{B_0}{4\sqrt{2}\pi^{3/2} G\bar{\rho}^{3/2}_{0} \left( v_{\mathrm{coll}}/2 + B_0/\sqrt{8\pi \bar{\rho}_0} \right)}}\nonumber \\
&\simeq &
 1.0\ \mathrm{Myr}
 \left(
\frac{B_0}{10\ \mathrm{\mu G}}
\right)^{1/2}
\left(
\frac{\bar{n}_{0}}{100\ \mathrm{cm^{-3}}}
\right)^{-3/4}\nonumber\\
&\times & 
\left[ 
\left(
\frac{v_{\mathrm{coll}}}{12\ \mathrm{km/s}}
\right)
+ 0.17
\left(
\frac{B_0}{10\ \mathrm{\mu G}}
\right)
\left(
\frac{\bar{n}_{0}}{100\ \mathrm{cm^{-3}}}
\right)^{-1/2}
\right]^{-1/2},
\label{equation:freefalltime}
\end{eqnarray}
where $\bar{\rho}_1 \simeq \sqrt{2}\mathcal{M}_{\mathrm{A}} \bar{\rho}_{\mathrm{0}}$ and $\bar{v}_{\mathrm{Alf}} = B_0/\sqrt{4\pi\bar{\rho}_0}$ are the mean density of the shocked layer and the mean Alfv\'{e}n velocity, respectively (from eq. [\ref{equation:shock jump cond}]), and $\bar{v}_{\mathrm{sh}} = v_{\mathrm{coll}}/2+v_1 \simeq v_{\mathrm{coll}}/2 + B_0/\sqrt{8\pi \bar{\rho}_0}$ represents the mean shock velocity (eq.~[\ref{equation:shock velocity}]).

Figure \ref{fig:tfilv12} represents the evolution of the filament mass fraction for the high shock velocity models, in which the time is normalized by the free-fall time $t_{\mathrm{ff}}$.
As we have shown in \S~\ref{subsec:fast coll case}, major filaments in these models are created by type O mechanism.
Figure \ref{fig:tfilv12} confirms that the formation timescale of the filaments by type O mechanism is much faster than the timescale of self-gravity.

Figure \ref{fig:tfilv3} shows the evolution of the filament mass fraction for the low shock velocity models.
In the results of model v3GyTn (panel c), the filament formation time $t_{\mathrm{fil}}$ coincides with the free-fall time $t_{\mathrm{ff}}\simeq 1.6\ \mathrm{Myr}$.
This supports our discussion in \S~\ref{subsec:slow coll case} such that the filaments are created by type G mechanism.
In panels (d) and (d'), type C occurs earlier than the free-fall time.

Figure \ref{fig:fraction_th} shows the evolution of the filament mass fraction (solid lines) and the filament formation time (dashed lines) for the several shock velocity models without turbulent velocity fluctuation such as v12GyTn (cyan), v10GyTn (blue), v8GyTn (purple), v6GyTn (red), and v3GyTn (orange).
We can confirm that the filament formation time increases with decrease in shock velocity, indicating that the dominant filament formation mechanism gradually changes with a decrease of shock velocity from type O to type G mechanism.
Figure \ref{fig:fraction_th} suggests threshold collision velocity for type O mechanism is approximately $v_{\mathrm{coll}}\sim 4.5\ \mathrm{km\ s^{-1}}$, corresponding to $v_{\mathrm{sh}}\sim 3.3\ \mathrm{km\ s^{-1}}$.


\subsection{Filament Line Mass Function} \label{subsec:Filament Line Mass Function}
We calculate the line-mass histogram based on the filament skeletons identified in the previous section.
For this purpose, we first determine the direction perpendicular to the filament at each grid on the filament skeleton in the projection plane, and we then compute the line-mass of the filament at each skeleton grid by integrating the gas column density along the normal directions within 0.1 pc from the skeleton grid.
Thus, the line-mass evaluated in our histogram is not the average line-mass of each filament but is the local apparent line-mass at each skeleton grid in the 2D projection plane.


In the top panels of Figure \ref{fig:lmfv12}, we plot the line-mass functions of model v12GyTn at $t \simeq t_{\mathrm{fil}}$ (top-left) and a later stage of $t=1.3\ \mathrm{Myr}$ (top-right).
The bottom panels are the same as the top panels but for model v12GyTy.
The horizontal and vertical axes are the local line-mass of the filament at a skeleton pixel (or a point on the filament axis) $M_{\mathrm{line}}$ and the number of pixels on filament axes (skeletons), respectively.
The pixel size equals to 6/512 pc $\simeq$ 0.012 pc.
The black line represents the power-law function with the Salpeter index, and the green and the black dashed lines show the critical line-mass with and without magnetic field support, respectively, i.e., $M_{\mathrm{line,cr}}=2c_{\mathrm{s}}^2 /G \sim 17\ \mathrm{M_{\odot}\ pc^{-1}}$~\protect\citep{Stodolkiewicz1963,Ostriker1964} and
\begin{eqnarray}
M_{\mathrm{line,cr,B}} 
&=& 2.24 \frac{B_1 w}{G^{1 / 2}}
+15.4 \frac{c_{s}^{2}}{G}
\nonumber \\
& \simeq& 13.5\left(\frac{w}{0.1\  \mathrm{pc}}\right)
\left(\frac{\bar{n}_{0}}{100\ \mathrm{cm^{-3}}}\right)^{1/2}\nonumber \\
& & \times 
\biggl{[}
\left(\frac{v_{\mathrm{coll}}}{12\ \mathrm{km\ s}^{-1}}\right)
\nonumber \\
& &+ 0.17 \left(\frac{B_{0}}{10\ \mu\mathrm{G}}\right)\left(\frac{\bar{n}_{0}}{100\ \mathrm{cm^{-3}}}\right)^{-1/2}
\biggr{]}
M_{\odot} \mathrm{pc}^{-1}\nonumber \\
& &+15.4\left(\frac{c_{s}}{0.2 \mathrm{km\ s}^{-1}}\right)^{2} 
M_{\odot} \mathrm{pc}^{-1},
\label{equation:line-mass with mag support}
\end{eqnarray}
derived by \protect\citet{Tomisaka2014}, where $B_1=\sqrt{2}\mathcal{M}_{\mathrm{A}}B_0$ and $w$ are the mean magnetic field in the shocked region and the width of filaments, respectively.
Note that eq.~(\ref{equation:line-mass with mag support}) is estimated by using the mean magnetic field strength in the shocked layer, which does not give the exact magnetic field strength threading the filaments.
In models v12GyTn and v12GyTy, most of the line-masses are sub-critical at $t \simeq t_{\mathrm{fil}}$ (panels [e] and [g] in Figure \ref{fig:lmfv12}), but they quickly evolve into super-critical ones by continuous accretion flows induced by the oblique shock in roughly 1 Myr (panels [f] and [h] in Figure \ref{fig:lmfv12}).


In panels (f) and (h) of Figure \ref{fig:lmfv12}, we can see that the filament line-mass functions have a Salpeter-like slope $\propto M_{\mathrm{line}}^{-1.35}$ at large line-masses.
This slope is similar to the line-mass function found in \protect\citet{Andre2019}.
Interestingly, this slope is the same as the high mass part of the core mass function of \protect\citet{Inutsuka2001}.
To understand the physical reason for this agreement between specific models and the Salpeter-like slope, we have to do more simulations by varying the parameters in our models.
This will be done in our next work.

Recent observations suggest that high line-mass filaments greater than $100\ \mathrm{M_{\odot}\ pc^{-1}}$ are strong candidates for massive star progenitors~\protect\citep{Fukui2019,Shimajiri2019,Tokuda2019II}.
We stress that such high line-mass filaments are naturally created in high shock velocity models in a short time.

In Figure \ref{fig:lmfv3}, we show the filament line-mass function for the low shock velocity cases at $t\simeq t_{\mathrm{fil}}$ (left panels) and at the time when the maximum line-mass exceeds $100\ \mathrm{M_{\odot}\ pc^{-1}}$ (right panels).
In the results of model v3GyTn, the filaments have almost critical line-mass at $t=t_{\mathrm{fil}}$ (panel i), which is consistent with our finding that the filaments are formed by type G mechanism in this model.
Panels (k) and (l) for model v3GyTy show that type C mode can create super-critical filaments.
However, the formation of massive filaments around $100\ \mathrm{M_{\odot}\ pc^{-1}}$ requires a relatively long time, i.e., more than 2 Myr after $t=t_{\mathrm{fil}}$.


To compare the histograms in models without self-gravity to the ones with self-gravity, in Figure \ref{fig:lmf_nog}, we show line-mass functions for models v12GnTn, v12GnTy, v3GnTn, and v3GnTy.
The line-mass functions in Panels (m)-(p) are similar to the ones in panels (e)-(h), respectively. 
This indicates that star-forming filaments can be created regardless of self-gravity.
In panel (t), we can see that low line-mass filaments disappear, and a massive part of filament mass function is truncated comparing to panel (l).
This indicates that low line-mass filaments created by type C are transient and self-gravity is needed to form massive filaments.
It should be noted that we impose a column density threshold of $1.5\times \bar{N}_{\mathrm{sh}}$ for the filament identification.
The threshold column density for the filament identification in model v3GnTy at $t$ = 2.3 Myr is set at $6.5\times 10^{21}\ \mathrm{cm^{-2}}$.
The corresponding minimum line-mass of the identified filament is $13\ \mathrm{M_{\odot}\ pc^{-1}}$ when the filament width is 0.1 pc.

\subsection{Role of Shear: Angle between Filaments and Magnetic Field} \label{subsec:Role of Shear}

In type S mechanism, shear motions in turbulence create faint filaments parallel to the magnetic field~\citep{SolerHennebelle2013,SolerHennebelle2017,KortgenSoler2020,InoueInutsuka2016}.
To study the role of the shear flow in more depth, we calculate probability distribution histograms as a function of angles between the filaments and the mass-weighted average magnetic field in the projection plane.
Since we already know the filament skeletons, it is straightforward to compute the angle.

We show angle histograms of model v12GyTy at $t$~=~0.2 (panels a and b),  0.4 (panels c and d), and 0.8 (panels e and f) Myr in Figure \ref{fig:angle_v12}.
The top and bottom panels are results when $t\simeq t_{\mathrm{fil}}$~(see \S\ref{subsec:Filament Formation Timescale} and Figure \ref{fig:tfilv12} panel b) and $t\simeq t_{\mathrm{ff}}$~(eq. [\ref{equation:freefalltime}]), respectively.
Left panels (a), (c), and (e) are the angle histograms for the filaments in the column density range of $0.5\ \bar{N}_{\mathrm{sh}}$ to $1.5\ \bar{N}_{\mathrm{sh}}$.
Right panels (b), (d), and (f) are the histograms for the filaments with $N>1.5\ \bar{N}_{\mathrm{sh}}$.
We see that most filaments, in particular dense filaments, in the right panels are perpendicular to the magnetic field.
In the later times, the fainter filaments shown in the left panels change the angle from perpendicular to parallel to the local magnetic field.
The effect of turbulent shear (type S mode) can naturally account for this evolution.

In Figure \ref{fig:angle_v3}, we show angle histograms of model v3GyTy at $t$ = 0.4 (panels g and h), 1.15 (panels i and j) and 2.0 (panels k and l) Myr.
Top and bottom panels are results at $t\sim t_{\mathrm{fil}}$ and $t \sim t_{\mathrm{ff}}$, respectively.
As Figure \ref{fig:angle_v12}, the left panels are for filaments having the column density in the range of $0.5\ \bar{N}_{\mathrm{sh}}$ to $1.5\ \bar{N}_{\mathrm{sh}}$, and the right panels are for filaments with $1.5\ \bar{N}_{\mathrm{sh}}$.
The trend is basically the same as the result of larger shock velocity simulation (v12GyTy; Figure \ref{fig:angle_v12}), but the angle distributions are more dispersed especially for fainter filaments (left panels).
This is because model v3GyTy has a slower evolution timescale than model v12GyTy and has more time for the fainter filaments to be stretched by the turbulent shear flows.

These results confirm that, except in the high column density region, the turbulent shear flow stretches the gas structure over time to form a low-density filamentary structure, but in this case, the orientation of the filaments is parallel to the magnetic field lines and the line-masses remain smaller than the critical line-mass~\citep{SolerHennebelle2013,SolerHennebelle2017,KortgenSoler2020,InoueInutsuka2016}.

\section{Discussion} \label{sec:Discussion}
\begin{figure}
    \begin{center}
        \includegraphics[clip,width=8.5cm]{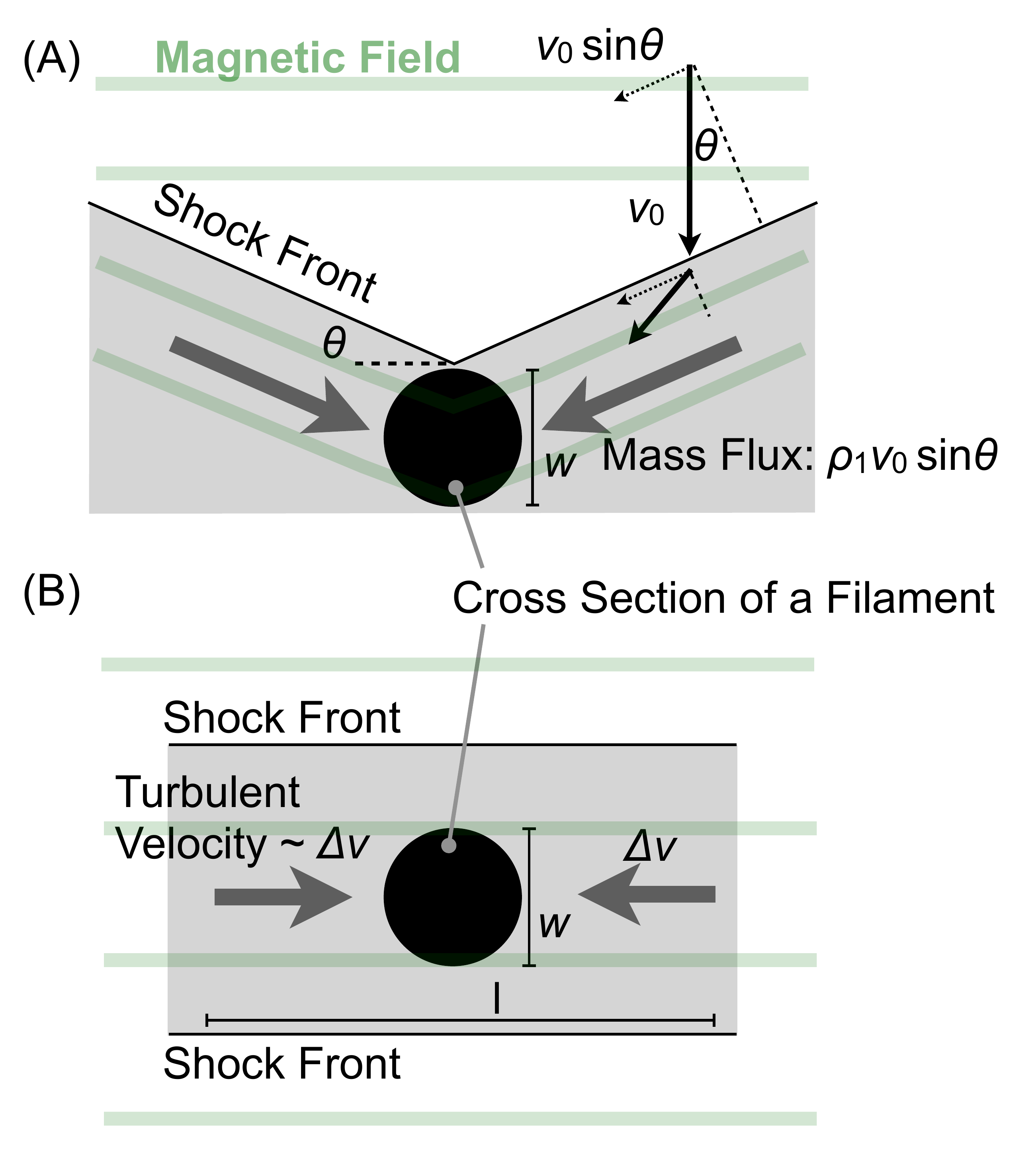}
        \caption{\small{Illustrations of the models used to estimate the filament formation timescale.
        The gray and black regions represent the post-shock layer and filament, respectively.
        (A): Schematic of the oblique MHD shock compression mechanism (type O), where $w$ is the filament width, and $\theta$ is the oblique shock angle.
        The angle depends on the detail of the interaction between the shock and the gas clump that evolves into the filament and is roughly $\theta \sim 30^{\circ}$ as indicated in Figures \ref{fig:pvv12Tn} and \ref{fig:pvv12Ty}.
        (B): Schematic of type C (compressive flows involved in initially given turbulence) mechanism, where $l$ is the scale of the turbulent compressive flow related to the velocity of the flow $\Delta v$ via Larson's law. Here we assume the width of the compressive flows as $w=0.1$ pc.
        }}
        \label{fig:model}
    \end{center}
\end{figure}

In this section, we estimate the growth timescale of the filaments formed by type O (oblique shock effect) and type C (compressive flows involved in initially given turbulence) mechanisms, which explain why the dominant filament formation mechanism changes with shock velocity.
The result given below emphasizes that type O and C mechanisms are distinct mechanisms having different characteristic timescales.
First, we estimate the filament formation timescale by type O mechanism.
As shown schematically in panel (A) of Figure \ref{fig:model}, the oblique MHD shock model predicts that the mass flux of the post-shock gas flow to the filament is given by $\rho_1 v_0 \sin{\theta}$, where $\theta$ is the oblique shock angle, and we have used the fact that the velocity component perpendicular to the shock normal is almost conserved across the shock\footnote{The conservation of the parallel velocity component is exact only for the limit of no magnetic field.
In the case of fast shock, the parallel velocity conservation is a highly accurate approximation.}.
If we assume constant width $w$ of the filament, the line-mass of the filament after time $t$ can be written as
\begin{equation}
    M_{\mathrm{line}} = 2{\rho}_{\mathrm{1}} v_{\mathrm{0}}w t \sin{\theta}.
\end{equation}
By substituting the isothermal strong MHD shock jump condition ${\rho}_{\mathrm{1}} \simeq \sqrt{2}\mathcal{M}_{\mathrm{A}}{\rho}_0$ (eq. [\ref{equation:shock jump cond}]), the timescale $t_{\mathrm{O}}$, in which type O mechanism creates filament of the line-mass $M_{\mathrm{line}}$, is estimated in the following equation.
\begin{eqnarray}
t_{\mathrm{O}} 
&=& \frac{M_{\mathrm{line}}}{2{\rho}_{\mathrm{1}} v_{\mathrm{0}} \sin{\theta} w} \nonumber \\
&=& \frac{M_{\mathrm{line}} B_0}{4\sqrt{2\pi}{\rho}^{3/2}_0 v^2_{\mathrm{0}}w \sin{\theta}} \nonumber \\
&=& 0.3\ \mathrm{Myr}\ \left(\sin{\theta}\right)^{-1}\ \left(\frac{v_{\mathrm{sh}}}{7\ \mathrm{km\ s^{-1}}}\right)^{-2} \nonumber \\
& \times & \left( 
    \frac{M_{\mathrm{line}}}{M_{\mathrm{line,cr}}} 
\right)
\left(
    \frac{{n}_{0}}{100\ \mathrm{cm^{-3}}} 
\right)^{-3/2}
\left(
    \frac{B_{0}}{10\ \mathrm{\mu G}} 
\right)
\left(
    \frac{w}{0.1\ \mathrm{pc}} 
\right)^{-1},
\label{equation:Obl time}
\end{eqnarray}
where we have used the fact that $v_0=v_{\mathrm{sh}}$.
Next, we consider the timescale of type C filament formation.
As shown in panel (B) in Figure \ref{fig:model}, the filament is formed when a turbulent flow converges in the shock-compressed slab.
In the following equation, we write the velocity of the converging flows in turbulence as $\Delta v$, their scale as $l$, and the width of the flow as $h$.
Because the origin of the converging flow is turbulence, we can use Larson's law $\Delta v \sim V_{08}\left(l / L_1 \right)^{0.5}$, where $V_{08}=0.8\ \mathrm{km\ s^{-1}},\ $and $L_1=1\ \mathrm{pc}$~\protect\citep{Larson1981}.
Then, the mass accumulation timescale is given by 
\begin{equation}
    t_{\mathrm{C}} = l/\Delta v \sim l^{1/2}L^{1/2}_1 / V_{08}.
    \label{mass accumulation timescale}
\end{equation}
By using $l$ and $w$, we can estimate the line mass of the filament as
\begin{equation}
    M_{\mathrm{line}} = {\rho}_{\mathrm{1}} l w.
    \label{the width of the flow w}
\end{equation}
By combining, eqs (\ref{mass accumulation timescale})-(\ref{the width of the flow w}), the timescale $t_{\mathrm{C}}$, in which the turbulent flow creates filament of the line-mass $M_{\mathrm{line}}$, can be estimated as
\begin{eqnarray}
    t_{\mathrm{C}} &=&
    \frac{M_{\mathrm{line}}}{\rho_1 w \Delta v} \nonumber \\
    &=&
    \left( \frac{M_{\mathrm{line}}L_1 B_0}{2\sqrt{2\pi}w V^2_{08} \rho_0^{3/2} v_{\mathrm{sh}}} \right)^{1/2} \nonumber \\
    &=& 0.78\ \mathrm{Myr}\ \left(\frac{v_{\mathrm{sh}}}{7\ \mathrm{km\ s^{-1}}}\right)^{-1/2} \left( 
        \frac{M_{\mathrm{line}}}{M_{\mathrm{line,cr}}} 
    \right)^{1/2}\nonumber \\
    &\times & 
    \left( 
        \frac{w}{0.1\ \mathrm{pc}} 
    \right)^{-1/2}
    \left(
        \frac{\bar{n}_{0}}{100\ \mathrm{cm^{-3}}} 
    \right)^{-3/4}
    \left(
        \frac{{B}_{0}}{10\ \mathrm{\mu G}} 
    \right)^{1/2}.
    \label{equation:eddy turn over time}
\end{eqnarray}
\begin{figure}
    \begin{center}
    \includegraphics[clip,width=8.5cm]{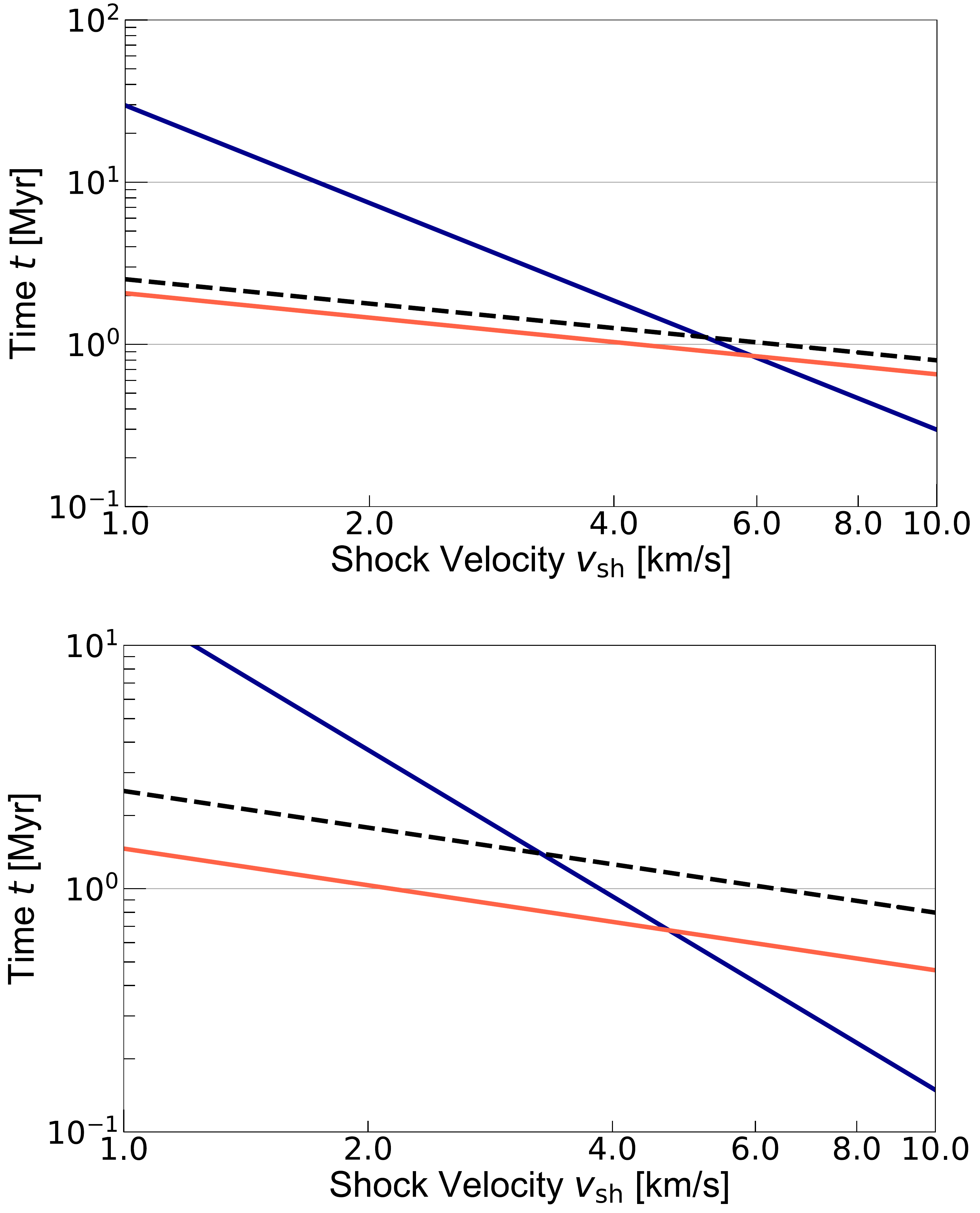}
        \caption{\small{Filament formation timescales as functions of shock velocity.
        The blue line shows the timescale for type O mechanism given by eq. (\ref{equation:Obl time}), and the red line shows type C timescale estimated by eq. (\ref{equation:eddy turn over time}).
        The dashed line represents the free-fall time in the shocked region $t_{\mathrm{ff}}$~(eq. [\ref{equation:freefalltime}]).
        \textit{Top panel}: Timescales required to reach the thermal critical line mass~\protect\citep[e.g.,][]{Stodolkiewicz1963,Ostriker1964}, i.e., $M_{\mathrm{line}}(t)= M_{\mathrm{line,cr}}$.
        \textit{Bottom panel}: Timescales required to reach half the thermal critical line mass, i.e., $M_{\mathrm{line}}(t)= M_{\mathrm{line,cr}}/2$.
        }}
        \label{fig:formationtime}
    \end{center}
\end{figure}

In the top panel of Figure \ref{fig:formationtime}, we plot the two timescales $t_{\mathrm{O}}$ and $t_{\mathrm{C}}$ for the critical line-mass filament ($M_{\mathrm{line}}=M_{\mathrm{line,cr}}$) as a function of the shock velocity.
We also plot the timescales for the filament with half the critical line-mass ($M_{\mathrm{line}} = M_{\mathrm{line,cr}}/2$) in the bottom panel.
The blue lines represent $t_{\mathrm{O}}$ with $\theta = 30^{\circ}$; the red lines are $t_{\mathrm{C}}$; and dashed lines are the free-fall time in the post-shock layer~(eq. [\ref{equation:freefalltime}]). 
When the shock velocity is high ($v_{\mathrm{sh}} \gtrsim 5\ \mathrm{km\ s^{-1}}$, i.e., $v_{\mathrm{coll}} \gtrsim 8\ \mathrm{km\ s^{-1}}$), we obtain the relationship of $t_{\mathrm{O}} < t_{\mathrm{C}} < t_{\mathrm{ff}}$, and for the lower shock velocity cases ($v_{\mathrm{sh}} \lesssim 5\ \mathrm{km\ s^{-1}}$, i.e., $v_{\mathrm{coll}} \lesssim 8\ \mathrm{km\ s^{-1}}$), we get $t_{\mathrm{C}} < t_{\mathrm{ff}} < t_{\mathrm{O}}$.
These results are fairly consistent with the results of the simulations.
That is, type O mechanism is important for the high shock velocity case, whereas type C is more effective for the low shock velocity case.


\section{Summary} \label{sec:Summary and Conclusion}
We have performed isothermal MHD simulations of the filament formation triggered by shock compression of a molecular cloud.
We found that when the shock is fast ($v_{\mathrm{sh}} \simeq 7\ \mathrm{km\ s^{-1}}$), the oblique MHD shock induced flows (type O mechanism defined in \S 1) works as the major mechanism for the formation of star-forming filaments irrespective of the presence of initial turbulence and self-gravity.
When the shock is slow ($v_{\mathrm{sh}} \simeq 2.5\ \mathrm{km\ s^{-1}}$), compressive flows involved in supersonic turbulence induce transient filament formation (type C), but the resulting filaments disperse unless the line-masses are comparable or larger than the thermal critical line-mass.
If we initially input strong turbulence with velocity dispersion larger than $\sim$ 5 km s$^{-1}$ in the simulation, shock waves locally occur in the simulation.
Thus, in principle, the type O process may occur in such simulations.
We cannot call type O ``turbulent" filamentation.
On the other hand, type O can be almost always induced by any single compression of a molecular cloud by interaction with relatively fast large-scale shock waves, e.g., by an expanding HII region, a supernova remnant, or a super-shell.
Note that type O and type C are observationally distinguishable through a characteristic structure in the position-velocity map and/or curve magnetic fields \citep[see,][]{Arzoumanian2018,Tahani2018,Tahani2019,Chen2020,Kandori2020a,Kandori2020b}.
When the shock velocity is low and no turbulence is set initially, the fragmentation of the shock-compressed sheet by self-gravity creates filaments (type G) over the gravitational fragmentation timescale of the dense sheet.
Formation of filaments by turbulent sheet-sheet collision (type I) was not clearly observed in our simulations, because this mode seems to be activated only when we initially set a weakly magnetized, uniform density molecular cloud.

By developing simple analytical models, we have shown that for $v_{\rm sh}\gtrsim 5$ km s$^{-1}$, type O responsible for creating major filaments, while type C is more effective for $v_{\rm sh}\lesssim 5$ km s$^{-1}$.
We conclude that the dominant filament formation mode changes with the strength of the incident shock wave.
Moreover, we analyzed the line-mass distribution of the filaments and showed that strong shock waves can naturally create high-line-mass filaments such as those observed in the massive star-forming regions~($M_{\mathrm{line}} \gtrsim 100\ \mathrm{M_{\odot}\ pc^{-1}}$) in a short time.
We stress that such high line-mass filaments are naturally created in high shock velocity models in a timescale of the creation of dense compressed sheet-like region.

\section{Acknowledgments}
We thank K. Iwasaki and D. Arzoumanian for fruitful discussions. The numerical computations were carried out on XC50 system at the Center for Computational Astrophysics (CfCA) of National Astronomical Observatory of Japan. This work is supported by Grant-in-aids from the Ministry of Education, Culture, Sports, Science, and Technology (MEXT) of Japan (15K05039, 18H05436).

\bibliography{ms}{}
\bibliographystyle{aasjournal}

\appendix
\restartappendixnumbering

\section{High Shock Velocity Case with Strong Turbulence}
\begin{figure*}
    \begin{center}
        \includegraphics[clip,width=17.5cm]{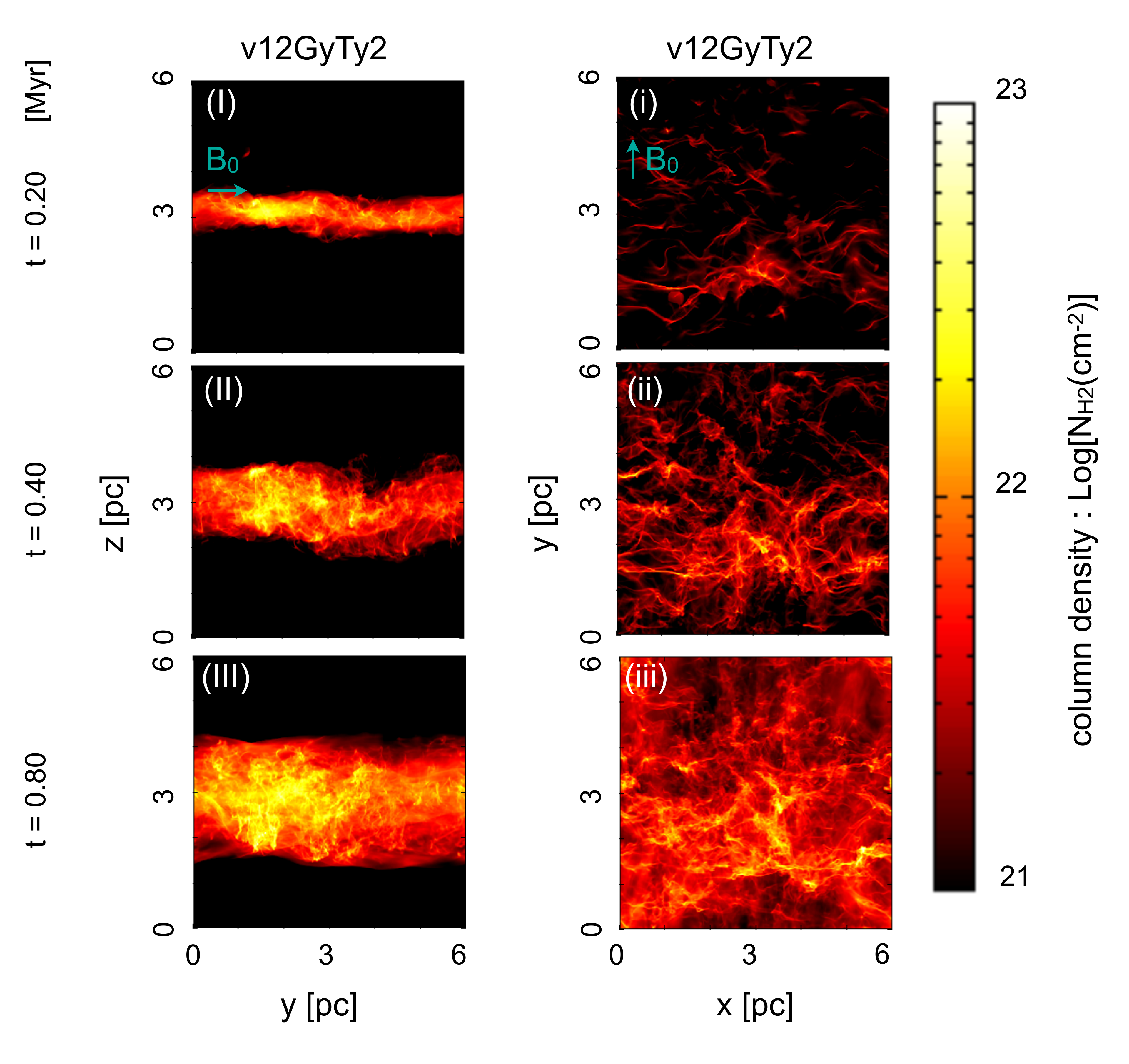}
        \caption{\small{Column density maps at time $t = 0.2\ (top),\ 0.4\ (middle),\ \mathrm{and}\ 0.8\ (bottom)\ \mathrm{Myr}$.
        \textit{Left row }(panels I, II, and III): Column density in the $y$-$z$ plane of model v12GyTy2.
        \textit{Right row }(panels i, ii, and iii): Same as the panels (I)-(III) but for the $x$-$y$ plane.
        }}
        \label{fig:v12st}
    \end{center}
\end{figure*}
\begin{figure}[ht!]
    \begin{center}
        \includegraphics[clip,width=17.5cm]{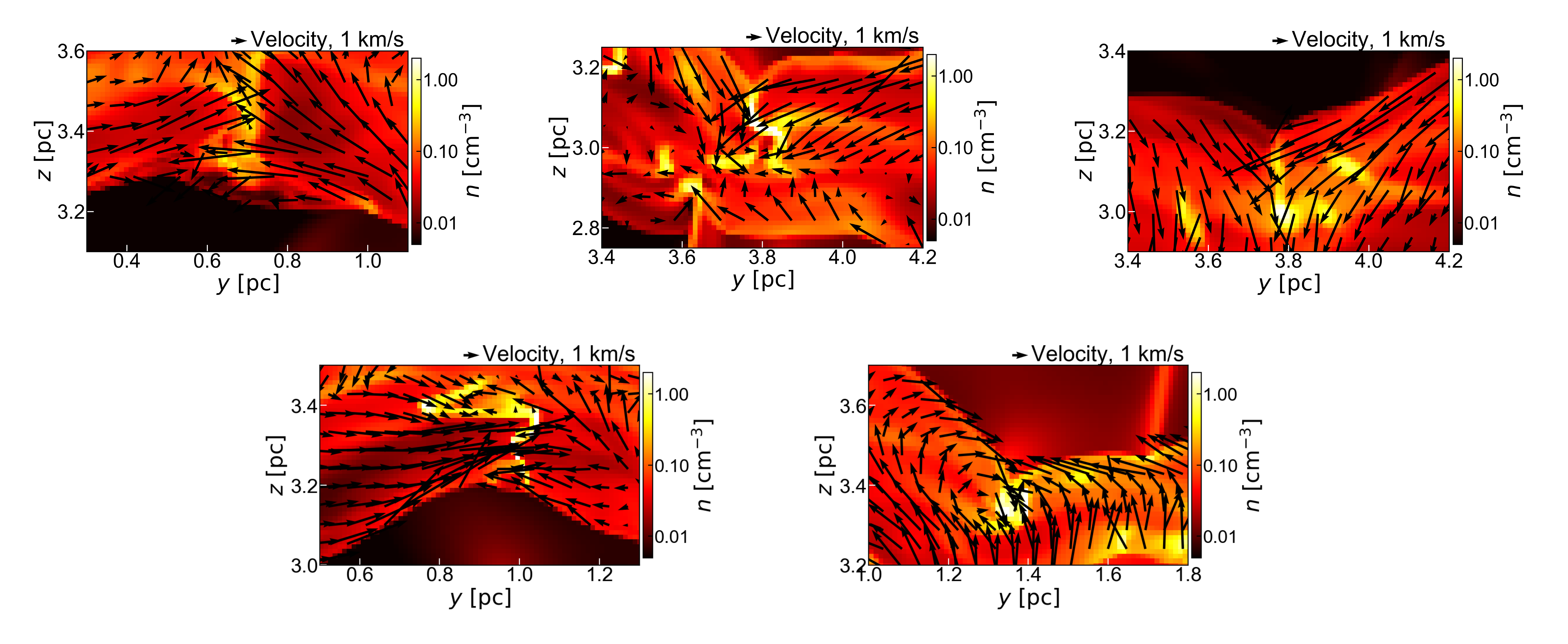}
        \caption{\small{Cross-section maps of the number density in the $y$-$z$ plane of model v12GyTy2.
        The yellow blobs located roughly at the center of each panel correspond to cross-sections of the filaments.
        We can confirm that the oblique MHD shock compression mechanism takes a major role in the filament formation even in the case of initial stronger turbulence with high shock velocity.
        }}
        \label{fig:v12stcrosssec}
    \end{center}
\end{figure}
\begin{figure*}
    \begin{center}
        \includegraphics[clip,
        width=18.05 cm]{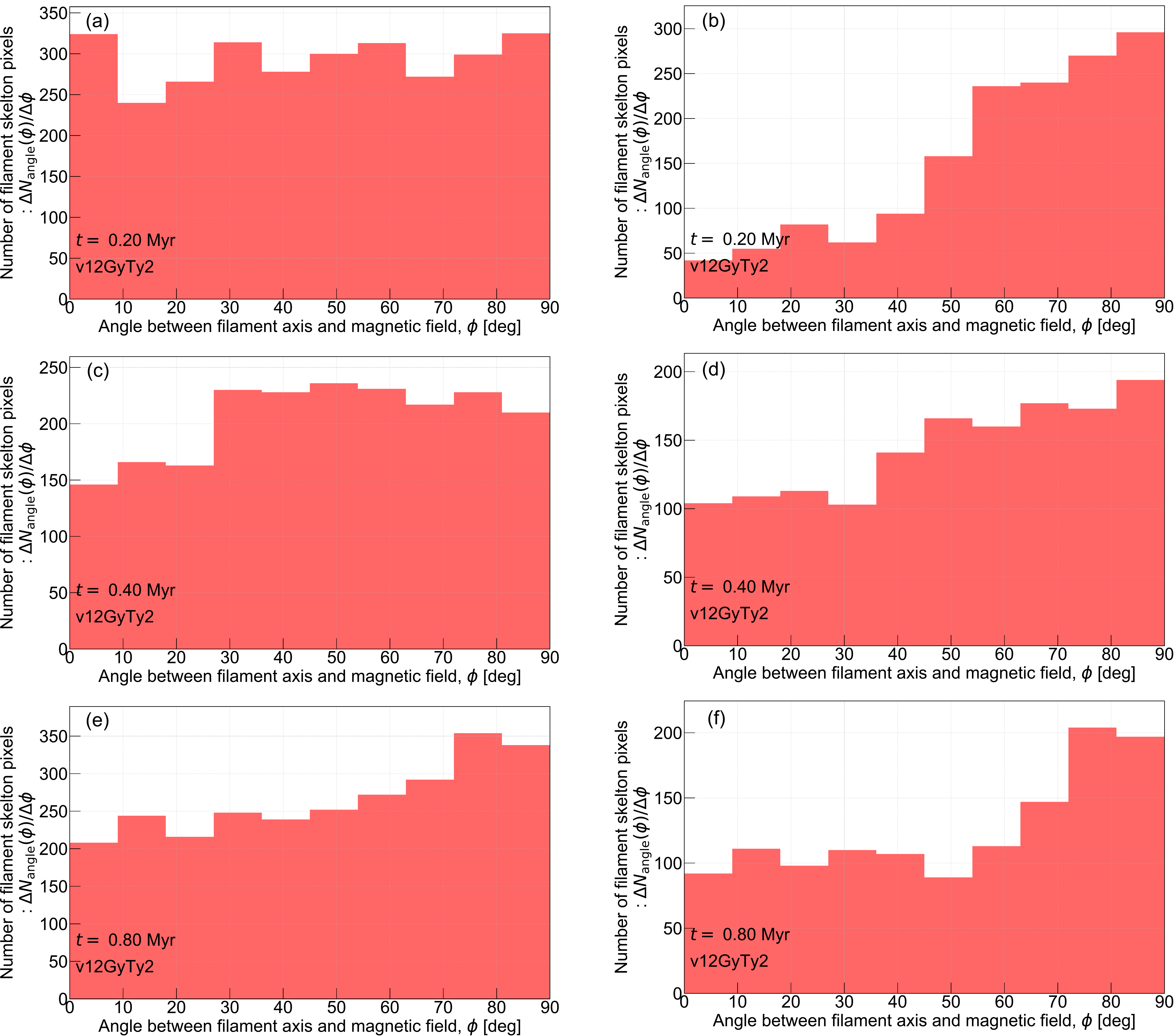}
        \caption{\small{Histogram of angles between filaments and magnetic field for model v12GyTy2.
        From top to bottom, results at time $t$=\ 0.2, 0.4 and 0.8 $\mathrm{Myr}$, respectively.
        \textit{Left panels} (1), (3), and (5): Results when we identify filaments in the column density range of 0.5$\bar{N}_{\mathrm{sh}}$ to 1.5$\bar{N}_{\mathrm{sh}}$.
        \textit{Right panels} (2), (4), and (6): Results when the filament identification threshold column density is chosen to be 1.5$\bar{N}_{\mathrm{sh}}$.
        }}
        \label{fig:angle_v12st}
    \end{center}
\end{figure*}

In the model v12GyTy, the velocity dispersion of the turbulence is set to $1~\mathrm{km\ s^{-1}}$.
In this appendix, we investigate the filament formation mechanism and the distribution of the angle between the filament and the magnetic field lines in the case of larger velocity dispersion of the initial turbulence $\Delta v\ =\ 2~\mathrm{km\ s^{-1}}$  as model `v12GyTy'.
Panels (I)-(III) and (i)-(iii) in Figure \ref{fig:v12st} show column density snapshots of model v12GyTy2 in the $y$-$z$ and $x$-$y$ planes at $t$~=~0.2~Myr, 0.4~Myr, and 0.8~Myr, respectively.
We can confirm that stronger turbulence make the structure more complex than the results in model v12GyTy. 
To clarify the dominant filament formation mechanism, we show the local density cross-sections around the five major filaments as the results of models v12GyTy2 in Figure \ref{fig:v12stcrosssec}.
The snapshots show that turbulent motion parallel to the magnetic field ($y$-direction) create dense clamps in the pre-shock region\footnote{This resembles type C filament formation, but this process happens before shocked sheet formation, and thus preshock clumps are formed.} that are swept by the shock and induces type O filament formations.
The curved shock morphology and velocity vectors (black arrows) shown in the cross-section panels in model v12GyTy2 support the activation of type O mechanism.

We show angle histograms of model v12GyTy2 at $t$~=~0.2 (panels a and b),  0.4 (panels c and d), and 0.8 (panels e and f) Myr in Figure \ref{fig:angle_v12st}.
Left panels (a), (c), and (e) are the angle histograms for the filaments in the column density range of $0.5\ \bar{N}_{\mathrm{sh}}$ to $1.5\ \bar{N}_{\mathrm{sh}}$.
Right panels (b), (d), and (f) are the histograms for the filaments with $N>1.5\ \bar{N}_{\mathrm{sh}}$.
While dense filaments are perpendicular to the magnetic field, the distribution of the angle between the faint filament and the magnetic field is random and different from that in Figure \ref{fig:angle_v12}.
It is remarkable that the histograms of the angle distribution is much flatter than the case of model v12GyTy (see, Fig.~\ref{fig:angle_v12st}).
This may be due to the fact that, when the initial turbulence is strong, various filamentation modes are mixed and the resulting filamentary structure changes its direction with respect to the magnetic field lines.
Observational study of density structure and magnetic field orientations in several molecular clouds by \citet{PlanckCollaboration2016} does not show such a flat distribution.
Therefore, we conclude that this kind of initial strong turbulence model does not provide a realistic model.

\section{High Shock Velocity Case without Magnetic Field}
\begin{figure*}
    \begin{center}
        \includegraphics[clip,width=17.5cm]{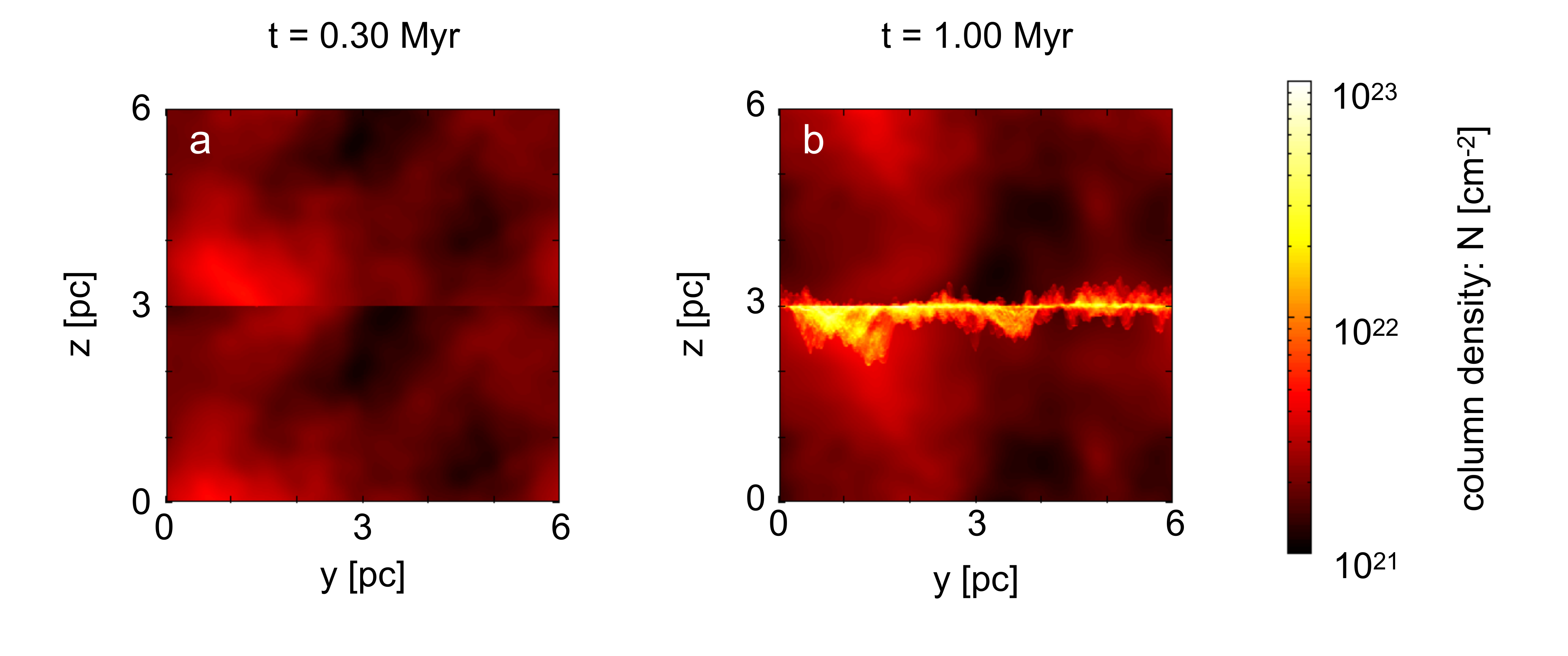}
        \caption{\small{Column density maps at time $t = 0.3\ (panel a),\ \mathrm{and}\ 1.0\ (panel b)\ \mathrm{Myr}$ in the $y$-$z$ plane of model v8GnTnB0.
        }}
        \label{fig:v8b0}
    \end{center}
\end{figure*}

In this section, we describe the additional set of calculations without magnetic field. 
The compression ratio of the isothermal hydrodynamics shock is given by $R \sim \mathcal{M}^2_{\mathrm{s}}$, where $\mathcal{M}_{\mathrm{s}}$ is the sonic Mach number.
Thus, for instance, if we set $v_{\rm sh}=7.0$ (or 2.5) km s$^{-1}$, the compression ratio become 1225 (or 156), which is extremely high.
Note that, with the magnetic field, the compression ratio is dramatically reduced to $R \sim \sqrt{2}\mathcal{M}_{\mathrm{A}}$, where $\mathcal{M}_{\mathrm{A}}$ is the Alfv\'{e}nic Mach number.

Here, we chose parameter $v_{\mathrm{coll}}\ \sim$~8.0~km~s$^{-1}$ ($v_{\mathrm{sh}}\ \sim$~5~km~s$^{-1}$), which we call v8GnTnB0.
Panels a and b in Figure \ref{fig:v8b0} show column density snapshots of model v8GnTnB0 in the $y$-$z$ at $t$~=~0.3~Myr, and 1.0~Myr, respectively.
In contrast to other column density figures in the main text, we integrate the whole numerical domain in constructing the column density.
In panel b, we see the highly unstable feature of shock surfaces.
This seems to be nonlinear thin-shell instability~\protect\citep[][]{Vishniac1994}, which is known to be suppressed in the case with a magnetic field.
A similar feature is observed in previous similar studies \protect\citep[e.g.,][]{Folini2014}.
In both panels, compressed layers sandwiched by two shocks are extremely thin and dense, which has not been observationally identified in the actual molecular clouds associated with fast shock waves as far as we are aware. 
Under this high  compression, the length scale of the initial fluctuations becomes smaller than the spatial resolution, and thus, the smallest scale structures shown in this section are limited by numerical resolution.  Nonetheless, we can identify no sign of the formation of the filamentary structure with sufficiently large line-mass. 
Therefore, we think that this setup without magnetic field is not realistic and the presence of magnetic fields is critical for creating filamentary structures in the shock compressed region.




\end{document}